\documentclass{egpubl}
\usepackage{eurovis2026}

\SpecialIssuePaper         % uncomment for final version of Computer Graphics Forum, special issue
\CGFccby
\usepackage[T1]{fontenc}
\usepackage{dfadobe}  

\BibtexOrBiblatex
\usepackage[backend=biber,bibstyle=EG,citestyle=alphabetic,backref=false]{biblatex} 
\electronicVersion
\PrintedOrElectronic
\ifpdf \usepackage[pdftex]{graphicx} \pdfcompresslevel=9
\else \usepackage[dvips]{graphicx} \fi

\usepackage{egweblnk}
\hypersetup{
}

\usepackage{amssymb}
\usepackage{array}
\usepackage{booktabs}
\usepackage{csquotes}
\usepackage{enumitem}
\usepackage{makecell}
\usepackage{mathtools}
\usepackage{multirow}
\usepackage{placeins}
\usepackage{subfig}
\usepackage[many]{tcolorbox}
\usepackage{todonotes}

\renewcommand{\quote}[1]{``\textit{#1}''}

\definecolor{blueDatasets}{HTML}{0096FF}
\definecolor{blueKnowledgeGraph}{HTML}{0433FF}
\definecolor{orangeSelection}{HTML}{FF9300}
\definecolor{purpleStaging}{HTML}{FF2F92}
\definecolor{greenSequencing}{HTML}{008F00}
\definecolor{yellowTransition}{HTML}{ABA95F}
\definecolor{grayScenes}{HTML}{919191}
\definecolor{greenDataTours}{HTML}{00FB92}
\definecolor{redSemanticTours}{HTML}{FF7E79}

\definecolor{main}{HTML}{1d4ed8}
\definecolor{sub}{HTML}{bfdbfe}
\newtcolorbox{boxCaseStudy}[1]{
    after skip = 0.5cm,
    attach title to upper={\ ---\ },
    before skip = 0.3cm,
    colback = sub,
    colframe = main,
    coltitle= main!75!black,
    boxrule = 1pt,
    boxsep = 0mm,
    left = 2mm,
    leftrule = 5pt,
    sharp corners,
    title = {\textbf{Case Study} #1}
}
\newtcbox{\entity}[3]{
    enhanced,
    nobeforeafter,
    tcbox raise base,
    boxrule=0.4pt,
    top=0mm,
    bottom=0mm,
    right=0mm,
    left=#2mm,
    arc=1pt,
    boxsep=2pt,
    colframe=#1!50!black,
    coltext=#1!25!black,
    colback=#1!10!white,
    overlay={
        \begin{tcbclipinterior}
            \fill[#1!75!blue!50!white] (frame.south west) rectangle node[text=white,font=\sffamily\bfseries\tiny] {#3} ([xshift=#2mm]frame.north west);
        \end{tcbclipinterior}
    }
}

\definecolor{ColorTaskIconText}{HTML}{0F203D}
\definecolor{ColorTaskIconBG}{HTML}{EBF0FA}
\definecolor{ColorTaskFrame}{HTML}{ADC2EB}
\newtcbox{\boxTask}{
    arc = 1pt,
    bottom = 1pt,
    boxrule = 0pt,
    boxsep = 0pt,
    colframe = ColorTaskFrame,
    colback = ColorTaskIconBG,
    coltext = ColorTaskIconText,
    left = 1.5pt,
    on line,
    right = 1.0pt,
    top = 1pt,
}
\definecolor{ColorInterpretationIconText}{HTML}{873b00}
\definecolor{ColorInterpretationIconBG}{HTML}{ffead9}
\definecolor{ColorInterpretationFrame}{HTML}{ADC2EB}
\newtcbox{\circleInterpretation}{
    bottom = 1.5pt,
    boxrule = 0pt,
    boxsep = 0pt,
    circular arc,
    colframe = ColorTaskFrame,
    colback = ColorInterpretationIconBG,
    coltext = ColorInterpretationIconText,
    left = 2pt,
    on line,
    right = 1.5pt,
    top = 1.5pt,
}

\newcommand{\framework}{SemanticTours}
\newcommand{\dataToursFramework}{DataTours}
\newcommand{\expert}[1]{$E_{#1}$}
\newcommand{\operator}[1]{$O_\mathrm{#1}$}
\newcommand{\selectionOperator}{\operator{sel}}
\newcommand{\stagingOperator}{\operator{sta}}
\newcommand{\sequencingOperator}{\operator{seq}}
\newcommand{\transitionOperator}{\operator{tra}}
\newcommand{\interviewQuestion}[1]{$IQ_{#1}$}
\newcommand{\researchQuestion}[1]{$RQ_{#1}$}
\newcommand{\interpretation}[1]{\circleInterpretation{$I_{#1}$}}
\newcommand{\interpretationGrammatical}{\interpretation{G}}
\newcommand{\interpretationTeleological}{\interpretation{T}}
\newcommand{\interpretationSystematic}{\interpretation{S}}
\newcommand{\interpretationHistorical}{\interpretation{H}}
\newcommand{\task}[1]{\boxTask{$T_{#1}$}}
\newcommand{\taskFactExtraction}{\task{1}}
\newcommand{\taskCounterfactuals}{\task{2}}
\newcommand{\taskDoctrinalization}{\task{3}}
\newcommand{\taskIssueIdentification}{\task{4}}
\newcommand{\taskConflictResolution}{\task{5}}
\newcommand{\taskNormInterpretation}{\task{6}}
\newcommand{\taskArgumentConstruction}{\task{7}}
\newcommand{\taskRuleConclusion}{\task{8}}
\newcommand{\taskAcademicCritique}{\task{9}}
\newcommand{\taskQuestion}[1]{$TQ_{#1}$}
\newcommand{\questionTaskFactExtraction}{\taskQuestion{1}}
\newcommand{\questionTaskCounterfactuals}{\taskQuestion{2}}
\newcommand{\questionTaskDoctrinalization}{\taskQuestion{3}}
\newcommand{\questionTaskIssueIdentification}{\taskQuestion{4}}
\newcommand{\questionTaskConflictResolution}{\taskQuestion{5}}
\newcommand{\questionTaskNormInterpretation}{\taskQuestion{6}}
\newcommand{\questionTaskArgumentConstruction}{\taskQuestion{7}}
\newcommand{\questionTaskRuleConclusion}{\taskQuestion{8}}
\newcommand{\ourToDo}[1]{\textcolor{red}{\textbf{#1}}}

\usepackage{xcolor}
\usepackage{colortbl}
\usepackage{siunitx}
\usepackage{tabularx}
\usepackage{makecell}
\newcommand\technique[1]{\mbox{\textbf{\textsc{#1}}}}

\definecolor{colorRay}{rgb}{0.4976, 0.6294, 0.871}
\definecolor{colorTouch}{rgb}{0.9533, 0.6651, 0.5031}
\definecolor{colorFilter}{rgb}{0.591, 0.86, 0.5745}
\definecolor{colorVolume}{rgb}{0.8875, 0.5608, 0.5608}
\definecolor{colorNeighborhood}{rgb}{0.709, 0.5965, 0.7941}
\definecolor{colorFishEye}{rgb}{0.7447, 0.5617, 0.4043}
\definecolor{colorGood}{RGB}{0,163,42}
\definecolor{colorBad}{RGB}{255,59,48}

\newcommand{\ray}{\textcolor{colorRay}{$\blacksquare$}~\technique{Ray}}
\newcommand{\touch}{\textcolor{colorTouch}{$\blacksquare$}~\technique{Touch}}
\newcommand{\filter}{\textcolor{colorFilter}{$\blacksquare$}~\technique{Filter}}
\newcommand{\volumee}{\textcolor{colorVolume}{$\blacksquare$}~\technique{Volume}}
\newcommand{\neighborhood}{\textcolor{colorNeighborhood}{$\blacksquare$}~\technique{Neighborhood}}
\newcommand{\fisheye}{\textcolor{colorFishEye}{$\blacksquare$}~\technique{Fisheye}}

\newcommand{\timecell}{\cellcolor{magenta!10}}
\newcommand{\physicalcell}{\cellcolor{cyan!10}}
\newcommand{\blockedcell}{\cellcolor{darkgray!20}}
\newcommand{\cellwidth}{0.955cm}
\newcommand{\cellwidthsmall}{0.66cm}

\newcommand{\colorbackground}[2]{%
  \begingroup
  \setlength{\fboxsep}{0pt}%
  \colorbox{#1}{#2}%
  \endgroup
}

\def\specialcell#1{\begin{tabular}[@{}\cellwidth@{}]{@{}r@{}}#1\end{tabular}}

\title[\framework{}]%
      {\framework{}: A Conceptual Framework for Non-Linear, Knowledge Graph-Driven Data Tours}

\author[Fürst et al.]
{
	\parbox{\textwidth}{\centering D. Fürst$^{1}$\orcid{0000-0002-0407-2867}, M. Jansen op de Haar$^{2}$\orcid{0009-0005-1673-592X}, M. El-Assady$^{3}$\orcid{0000-0001-8526-2613}, D. A. Keim$^{1}$\orcid{0000-0001-7966-9740}, and M. T. Fischer$^{1}$\orcid{0000-0001-8076-1376}}
	\\
	\parbox{\textwidth}{\centering $^1$University of Konstanz, Germany \\ $^2$University of Twente, Netherlands \\ $^3$ETH Zurich, Switzerland}
}
\begin{document}

% uncomment for using teaser
%\begin{comment}
\teaser{
    \centering
    \includegraphics[width=\linewidth]{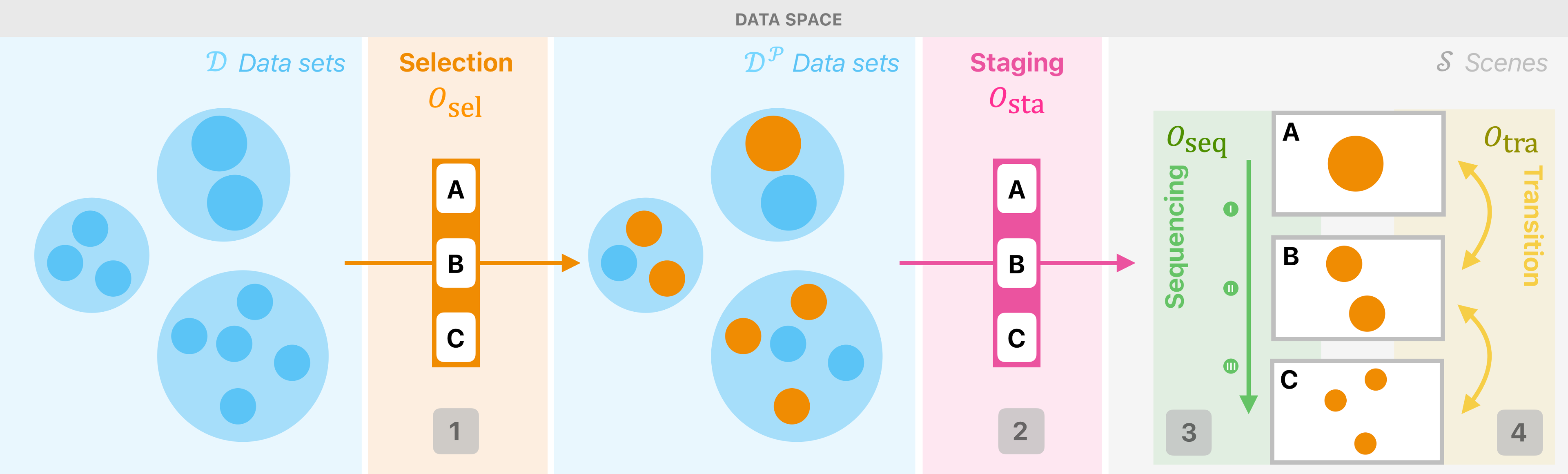}
    \caption{
        A conceptual overview of the data space of our model, \framework{}. Its operations enable non-linear, graph-driven tours for progressive exploration.
        The data space contains the data points of all available \textcolor{blueDatasets}{data sets}.
        The four operators (1)~\textcolor{orangeSelection}{select} relevant data points~(\selectionOperator{}), (2)~\textcolor{purpleStaging}{stage} them in a scene~(\stagingOperator{}), (3)~\textcolor{greenSequencing}{determine the order} of these \textcolor{grayScenes}{scenes}~(\sequencingOperator{}), and (4)~\textcolor{yellowTransition}{transition} between them~(\transitionOperator{}).
    }
    \label{figure:data-space}
}
%\end{comment}

\maketitle
%-------------------------------------------------------------------------
\begin{abstract}

% WHAT ARE TOURS?
Interactive tours help users explore datasets and provide onboarding.
They rely on a linear sequence of views, showing a curated set of relevant data selections and introduce user interfaces.
% WHAT IS THE RESEARCH GAP?
Existing frameworks of tours, however, often do not allow for branching and refining hypotheses outside of a rigid sequence, which is important in knowledge-centric domains such as law.
For example, lawyers performing analytical case analysis need to iteratively weigh up different legal norms and construct strings of arguments.
% WHAT DO WE CONTRIBUTE?
To address this gap, we propose \framework{}, a semantic, graph-based model of tours that shifts from a sequence-based towards a graph-based navigation.
% WHAT ARE THE DETAILS OF OUR CONTRIBUTION?
Our model constructs a domain-specific knowledge graph that connects data elements based on user-definable semantic relationships.
These relationships enable non-linear graph navigation that defines tours.
% WHAT DO WE CONTRIBUTE?
We apply \framework{} to the domain of law and conceptualize a visual analytics design and interaction concept for analytical reasoning in legal case analysis.
% WHAT ARE THE DETAILS OF OUR CONTRIBUTION?
Our concept accounts for the inherent complexity of graph-based tours using aggregated graph nodes and supporting navigation with a semantic lens.
During an evaluation with six domain experts from law, they suggest that graph-based tours better support their analytical reasoning than sequences.
% WHAT IS THE IMPACT OF OUR CONTRIBUTION?
Our work opens research opportunities for such tours to support analytical reasoning in law and other knowledge-centric domains.

\begin{CCSXML}
<ccs2012>
   <concept>
       <concept_id>10003120.10003121</concept_id>
       <concept_desc>Human-centered computing~Human computer interaction (HCI)</concept_desc>
       <concept_significance>300</concept_significance>
       </concept>
   <concept>
       <concept_id>10003120.10003145.10003147.10010365</concept_id>
       <concept_desc>Human-centered computing~Visual analytics</concept_desc>
       <concept_significance>500</concept_significance>
       </concept>
   <concept>
       <concept_id>10002951.10003317.10003331.10003271</concept_id>
       <concept_desc>Information systems~Personalization</concept_desc>
       <concept_significance>500</concept_significance>
       </concept>
   <concept>
       <concept_id>10002951.10003317.10003331.10003336</concept_id>
       <concept_desc>Information systems~Search interfaces</concept_desc>
       <concept_significance>500</concept_significance>
       </concept>
 </ccs2012>
\end{CCSXML}

\ccsdesc[300]{Human-centered computing~Human computer interaction (HCI)}
\ccsdesc[500]{Human-centered computing~Visual analytics}
\ccsdesc[500]{Information systems~Personalization}
\ccsdesc[500]{Information systems~Search interfaces}

\printccsdesc
\end{abstract}

%%%%%%%%%%%%%%%%%%%%%%%%%%%%%%%%%%%%%%%%%%%%%%%%%%%%%%%%%%%%%%%%%%%%%%%%%%%%%%%%%%%%%%%%%%%%%%%%

\section{Introduction}
Data tours present a way to guide users through datasets and interactive visualizations by presenting a curated sequence of views on data~\cite{asimovGrandTourTool1985, mehtaDataToursDataNarratives2017}. They, for example, have been used in data storytelling~\cite{mehtaDataToursDataNarratives2017}, introducing novices to datasets in event~\cite{yuAutomaticAnimationTimeVarying2010} or network analysis~\cite{liNetworkNarrativesDataTours2023}, supporting on-boarding~\cite{dhanoaDTourSemiAutomaticGeneration2024}, and enabling exploratory analysis across domains such as business intelligence~\cite{pererSystematicFlexibleDiscovery2008}, science~\cite{eckeltLoopsLeveragingProvenance2025}, and journalism~\cite{fuAtaEaverAuthoring2025}. By structuring a dataset into a linearly ordered path, tours help users orient themselves, discover relevant attributes, and important subsets~\cite{mehtaDataToursDataNarratives2017}, while building a progressive understanding of the data.

However, traditional tours primarily operate on predefined visual states of datasets. They are typically implemented as linear lists of data views, reflecting a fixed, attribute-driven perspective of the data and a slideshow-like navigation model~\cite{mehtaDataToursDataNarratives2017}. This approach is well-suited to showcasing existing insights, but it lacks the flexibility to adapt to complex analysis workflows, particularly those where hypotheses evolve dynamically~\cite{liNetworkNarrativesDataTours2023}. Moreover, these tours focus on data attributes in isolation, not accommodating higher-level domain semantics or the tacit knowledge of experts. As a result, theoretical frameworks such as DataTours~\cite{mehtaDataToursDataNarratives2017} and similar guidance systems remain constrained to static, view-centric narratives, struggling to support and present exploratory, hypothesis-driven work in knowledge-centric domains.

In many domains, especially in the social sciences, law, and medicine, analysis depends on abstract relationships that are not directly visible in the data~\cite{ruethersRechtstheorieUndJuristische2022}. 
Experts traverse concepts, relationships, and abstractions rather than merely views of a specific model configuration~\cite{liNetworkNarrativesDataTours2023}; they construct and refine mental models that connect data to evolving research questions. 
Supporting this kind of work requires moving beyond a fixed series of dataset views and towards an adaptive, semantic layer that captures relationships between concepts, encodes domain-specific knowledge, and integrates user-generated hypotheses into the exploration process.
In such settings, a tour is not just a linear path through data views but a tour through an interconnected network of related information, which can be represented as a knowledge graph~(KG).
Hence, we pose the following research questions:
\begin{itemize}[leftmargin=2.25em]
    % TODO: formulated to general? Maybe narrow down to our focus.
    \item[\researchQuestion{1}] How can knowledge graphs serve as a foundation for navigating non-linear analytical reasoning processes in knowledge-centric domains like law?
    % TODO: are these operations too specific? Did we miss some?
    \item[\researchQuestion{2}] What visual designs can support branching, detouring, and replaying analytical reasoning pathways?
\end{itemize}

In this paper, we contribute \emph{\framework{}}, a semantic, graph-based model for tours that operates over domain-specific entities and relations, while describing corresponding new non-linear navigation and interaction paradigms.
We achieve the semantic model by building a KG that captures important concepts and domain-specific relationships, relying on a user-definable, domain-specific knowledge transfer function acting as an extractor.
The KG links data-driven insights to domain concepts and their relationships, enabling users to traverse connected ideas.
By linking individual data attributes with higher-level concepts, users externalize their tacit knowledge and can reuse analytical pathways across sessions. 
The KG acts as a semantic layer and a transfer function, mapping heterogeneous datasets to domain-specific conceptual spaces while remaining generic across application areas.
Unlike traditional tours, \framework{} allows non-linear exploration, where users can branch, detour, and replay their analysis steps.
These exploration paths form semantic workflows that represent how users connect facts within the data. This facilitates the externalization of knowledge, touring, and sharing workflows to support ongoing analysis.

\begin{comment}
Building on this conceptual foundation, we propose a set of novel user interface and interaction concepts for navigating non-linear, semantic tours. Our design supports branching, loops, and context-sensitive traversal, enabling adaptive pathways that reflect the fluidity of expert reasoning. We combine principles from data guidance, facet-based exploration, and KG visualization to define a framework for semantic navigation. In doing so, we introduce design patterns for authoring and interacting with tours that are flexible enough to accommodate evolving analysis workflows while maintaining a cognitively aligned structure for exploration.
\end{comment}

We apply \framework{} to the domain of law and conceptualize a visual analytics~(VA) design for analytical reasoning in legal case analysis.
Law is one example of a knowledge-centric domain that relies on document navigation, tacit domain knowledge, and intricate, yet accurate, reasoning processes.
Using non-linear, semantic tours, our concept supports branching, detouring, and replaying to enable adaptive pathways that reflect the fluidity of expert reasoning.
Our concept addresses the inherent complexity of graph-based tours by utilizing aggregated graph nodes and providing navigation with a semantic lens.
Aggregated graph nodes jointly visualize the immediate neighborhood of adjacent nodes.
The semantic lens highlights the current step of the analytical reasoning process, presenting an excerpt from a semantic tour, while blurring other irrelevant steps.
% First, (1) we present a case study on exploring legal document collections, showcasing how semantic tours enable experts to map and navigate complex relationships between laws, cases, and concepts.
We conduct a controlled user study with six legal experts, evaluating \framework{} and our VA design concept.
The expert feedback suggests that graph-based tours better support their legal analytical reasoning than sequences.
% demonstrating significant reductions in mental workload and measurable improvements in task performance for interpretive work.
% Together, these studies highlight the benefits of embedding domain semantics into tour-based exploration and provide insights into the design of cognitively aligned, adaptive exploration tools.
Thereby, we make the following contributions:
\begin{itemize}
    \item \framework{}, a semantic, graph-based framework of tours for designing VA applications that support analytical reasoning.
    \item A VA design concept that showcases the use of \framework{} for an application that enables legal case analysis in law.
    \item An evaluation of \framework{} and the VA design concept in the form of a qualitative user study with six legal experts.
\end{itemize}
Our work opens research opportunities for non-linear, semantic tours to support analytical reasoning in law and other knowledge-centric domains.

%%%%%%%%%%%%%%%%%%%%%%%%%%%%%%%%%%%%%%%%%%%%%%%%%%%%%%%%%%%%%%%%%%%%%%%%%%%%%%%%%%%%%%%%%%%%%%%%

% TODO: I think we need to incorporate the definition of explicit and implicit/tacit knowledge somehow!

\section{Background \& Related Work}
We extend prior work on data tours in storytelling and data exploration to analytical reasoning.
Through this, \framework{} also intersects with research on faceted search, guidance, and provenance.

\subsection{Data Tours for Storytelling}
% Definition of data tours
Tours are a family of visualization techniques for guiding readers through a curated sequence of views.
\dataToursFramework{}~\cite{mehtaDataToursDataNarratives2017} formalized a tour as a structured, hierarchical sequence of views over attributes and subsets of data, emphasizing how such sequences can reveal inherent relationships in a dataset. %, while allowing limited interactivity within each view.
For this, the states of visualization components, data filters, and viewport positions are serialized into a linear script that viewers step through.
The general idea of storytelling can be distinguished on a continuum between author-driven and reader-driven approaches~\cite{segelNarrativeVisualizationTelling2010}.
For the former, one example of visual storytelling is Infographics~\cite{lankowInfographicsPowerVisual2012} that are common in online journalism.
% \todo{author-driven vs. reader-driven, infographis, next we come to comics, can we have a more canocical /main stream /most cited example first?}
% Storytelling: author-driven
% \todo{it remains unclear we jump from data tours to bachs comics here}
A similar approach is data comics that present readers with images that serve as visualizations of data and represent the perspectives of a data tour~\cite{bachEmergingGenreData2017}.
The sequential order of data comics forms a narrative structure, mirroring the sequence of the tour.
% With their linear and rigid progression, data comics exemplify \textbf{author-driven storytelling} with data tours.
Such approaches are also characterized by a prominent narrative structure and limited interactivity, if any.
% Therefore, these approaches are commonly used in online journalism, for example.
% Storytelling: reader-driven
% However, they represent only one extreme of a continuum that spans from author-driven to reader-driven storytelling.
On the other end of the spectrum, reader-driven storytelling is characterized by a lack of prescribed ordering but a surplus of interactivity, prioritizing human agency and creativity.
For example, DataWeaver enables users to author data narratives by connecting interactive visualizations and text, weaving them into coherent explanations~\cite{fuAtaEaverAuthoring2025}.
There, nodes on a canvas represent the views of a tour, either as a visualization or text, and edges connecting these nodes form a flow that represents the sequence of the tour, thereby creating a narrative structure.
In the context of network visualizations, \citeauthor{takahiraTangibleNetSynchronousNetwork2025} leverage tours to enable storytelling through interactions with physical objects in synchronous contexts, e.g., for teachers in a classroom~\cite{takahiraTangibleNetSynchronousNetwork2025}.

% Data exploration
While still considered a form of storytelling, reader-driven approaches also facilitate \textbf{data exploration}.
For example, CLUE couples interactive data analysis with storytelling to document insights, enabling users to transition from sense-making to narrative construction and back within a single application~\cite{gratzlVisualExplorationStorytelling2016}.
For a more proactive approach, DataSite detects salient features of a dataset and surfaces them as notifications in a feed-like timeline, effectively assembling a data story from algorithmic findings~\cite{cuiDataSiteProactiveVisual2019}.
Like CLUE and DataSite, many practical approaches are hybrid, combining techniques from author- and reader-driven storytelling~\cite{segelNarrativeVisualizationTelling2010}.
These hybrid approaches enable analytical reasoning through data exploration and documentation, which serve as essential building blocks for storytelling~\cite{pirolli2005sensemaking}.

\subsection{Data Tours for Analytical Reasoning}
% Definition of sense-making and analytical reasoning
In an iterative process, information foraging and sense-making complement each other to enable analytical reasoning~\cite{pirolli2005sensemaking}.
\citeauthor{pirolli2005sensemaking} formalize this loop as a model of sense-making that consists of a foraging loop and a sense-making loop~\cite{pirolli2005sensemaking}.
Together, these two loops can be interpreted as a broader theory of analytical reasoning~\cite{cook2005illuminating}.
\par{\textbf{Information Foraging}} During the foraging loop, an analyst searches data and extracts information relevant to their problem of analysis.
Since this task is prevalent in many domains, several paradigms have emerged to support information foraging.
% Foraging: faceted search
For example, \citeauthor{hearstFindingFlowWeb2002} introduced faceted search~\cite{hearstFindingFlowWeb2002}, an interface design that supports both directed search and exploratory browsing through structured facets, typically using a facet tree paired with a result list.
Progressively, further research has enriched the paradigm of faceted search by leveraging spatial encodings and graph structures. 
For instance, FacetScape uses a Voronoi-based visualization to improve the display and navigation of facets~\cite{seifertFacetScapeVisualizationExploring2014}.
Meanwhile, \citeauthor{heimFacetedVisualExploration2011} and \citeauthor{guoGRAFSGraphicalFaceted2023} further improved the efficacy of faceted search by leveraging semantic links between data items and KGs~\cite{heimFacetedVisualExploration2011, guoGRAFSGraphicalFaceted2023}.
% Foraging: data tours
However, faceted search is yet to be combined with tours.
Instead, Systematic Yet Flexible~(SYF) has emerged as a separate strand of research in response to growing data sets and increasingly complex exploration processes~\cite{pererSystematicFlexibleDiscovery2008}.
SYF provides an overview of all actions available during exploration in a sequential manner, but allows for actions to be performed in any order to provide flexibility.
More recently, SYF inspired NetworkNarratives~\cite{liNetworkNarrativesDataTours2023}, a simplified, linear interpretation of DataTours~\cite{mehtaDataToursDataNarratives2017} that guides novices in network analysis.
Both SYF and NetworkNarratives demonstrate how tours can facilitate information foraging by guiding the user.

% Foraging: guidance
A computer-assisted process that helps resolve knowledge gaps of users during interactive analysis is coined as \textbf{guidance}~\cite{cenedaCharacterizingGuidanceVisual2017}.
It enables systems to direct attention, suggest next steps, or propose alternative views to mitigate uncertainty.
% \citeauthor{cenedaCharacterizingGuidanceVisual2017} distinguish between degrees of guidance and types of gaps, ranging from orienting to prescribing guidance.
Tours are typically employed to provide onboarding to users of visualizations and applications in guidance~\cite{dhanoaDTourSemiAutomaticGeneration2024}.
For example, DashGuide captures an author-performed sequence of interactions and generates the corresponding content for onboarding users to a dashboard~\cite{hoqueDashGuideAuthoringInteractive2025a}.
There, tours replay the sequence of captured user interactions step-by-step to highlight key features of a dashboard for new users.
% Sense-making: provenance
\par{\textbf{Sense-Making}} By closing knowledge gaps like these through guidance, tours can prove useful during information foraging. %, but also for the sense-making loop.
However, they are also beneficial during sense-making, where analysts come up with a schema of information extracted during the foraging loop to form, validate, and refine hypotheses~\cite{pirolli2005sensemaking}.
This requires a retrospective analysis of the steps conducted during the information foraging process.
In research, these steps are formalized as \textbf{provenance}, which refers to the capture of and communication about the history of data, visualization, interaction, insight, and rationale~\cite{kaixuAnalyticProvenanceSensemaking2015, raganCharacterizingProvenanceVisualization2016, kaixuSurveyAnalysisUser2020}.
The need for externalized support during analytical reasoning has been recognized before~\cite{shrinivasanSupportingAnalyticalReasoning2008}.
\citeauthor{shrinivasanSupportingAnalyticalReasoning2008} proposed an information visualization framework specifically designed to support the analytical reasoning process through three integrated views: a data view for interactive visualization, a knowledge view for recording analysis artifacts, and a navigation view for capturing visualization states.
Their approach enables analysts to externalize mental models, link analysis artifacts to specific visualization states, and revisit previous reasoning steps.
However, their framework focuses on traditional visualization states rather than semantic relationships.
% \todo{more provenance literature? e.g. which has been described extensivley [a,b,c,]. In context of tours, GraphTrail...}
% Sense-making: data tours
% \todo{make clear you take about provenance for data tours only for these examples}
In the context of tours, GraphTrail visualizes steps of analytical provenance for network analysis as a graph on a canvas, allowing users to revisit steps and understand how specific insights emerged from sequences of actions and views~\cite{dunneGraphTrailAnalyzingLarge2012}.
There, a data tour represents a user-driven sequence of views that correspond to the steps of an exploration path, forming a graph of tours that depicts the process of network analysis.
Other systems employ tours in similar ways to review past states of data~\cite{eckeltLoopsLeveragingProvenance2025, yanKNowNEtGuidedHealth2025}, interaction~\cite{akhileshcamisettyEnhancingWebbasedAnalytics2019, stitzKnowledgePearlsProvenanceBasedVisualization2019}, visualization~\cite{bavoilVisTrailsEnablingInteractive2005, kimGraphScapeModelAutomated2017, stitzKnowledgePearlsProvenanceBasedVisualization2019, yanKNowNEtGuidedHealth2025}, and rationale~\cite{nguyenSensePathUnderstandingSensemaking2016, mathisenInsideInsightsIntegratingDataDriven2019}.
As the underlying user interactions remain manual~\cite{parkStoryFacetsDesignStudy2022, eckeltLoopsLeveragingProvenance2025, yanKNowNEtGuidedHealth2025}, inferring precise provenance from low-level interactions to elaborate on the analytical reasoning still remains a challenge, given the nature of cognitive processes~\cite{kaixuAnalyticProvenanceSensemaking2015}.

% TODO: should we include liWhereAreWe2024 (Understanding Data Storytelling Tools from the Perspective of Human-AI Collaboration) to provide an angle on the role of AI in the context of our work?
\subsection{Non-Manual Construction of Data Tours}
Tours often require manual construction, and automatically creating them remains a challenge~\cite{mehtaDataToursDataNarratives2017, liNetworkNarrativesDataTours2023}.
% Research gap
In the context of data exploration, there have been attempts to tackle this challenge~\cite{wongsuphasawatVoyagerExploratoryAnalysis2016, liKG4VisKnowledgeGraphBased2022, zhaoChartSeerInteractiveSteering2022, chenCalliopeNetAutomaticGeneration2023, shiSupportingGuidedExploratory2023}.
% For instance, \citeauthor{shiSupportingGuidedExploratory2023} automatically create a tour of next steps to perform during data exploration with \emph{Visail}~\cite{shiSupportingGuidedExploratory2023} by mimicking human behavior through reinforcement learning.
For instance, \citeauthor{wongsuphasawatVoyagerExploratoryAnalysis2016} support data exploration by automatically recommending visualizations through faceted browsing in Voyager~\cite{wongsuphasawatVoyagerExploratoryAnalysis2016}. This recommendation is based on statistical and perceptual measures.
On the other hand, \citeauthor{liKG4VisKnowledgeGraphBased2022} leverage a KG that represents relationships between data features, data columns, and visualizations~\cite{liKG4VisKnowledgeGraphBased2022}.
The authors implement KG4Vis based on embeddings of that KG to recommend visualizations for a given dataset according to pre-defined rules.
However, they do not present a corresponding user interface~(UI).
Still, these approaches (1) remain linear, which does not accurately reflect human thought processes, (2) disregard semantic relationships of the domain, or (3) cover only selected steps of the analytical reasoning process without a holistic perspective.
% While some approaches, such as \emph{StoryFacets}~\cite{parkStoryFacetsDesignStudy2022}, closely align with the sense-making process, they necessitate manual construction of data tours.
% Conversely, approaches that automatically assemble data tours, like \emph{Visail}~\cite{shiSupportingGuidedExploratory2023}, remain linear and cover only selected steps of the analytical reasoning process.
To preserve human agency and creativity in analytical reasoning, automatically generating tours should not replace human-driven exploration or sense-making~\cite{fuAtaEaverAuthoring2025}.
Automatically generated tours contribute to a mixed-initiative environment, where the underlying system and the human analyst collaborate in reasoning~\cite {horvitzPrinciplesMixedInitiativeUser1999}.
Hence, automatically creating non-linear tours remains an underexplored challenge~\cite{mehtaDataToursDataNarratives2017, liNetworkNarrativesDataTours2023}.
In particular, combining the advantages of semantically linked data with the benefits of tours has received limited attention.

%%%%%%%%%%%%%%%%%%%%%%%%%%%%%%%%%%%%%%%%%%%%%%%%%%%%%%%%%%%%%%%%%%%%%%%%%%%%%%%%%%%%%%%%%%%%%%%%

% Manuel: problem/scenario, where the linear nature of data tours is limiting
%         extend the scenario from NetworkNarratives?
%         showcase jump backs and alternatives
%         formulas are hard to read/understand
%         graphics for the case study/operators
%         schematic visualization/figure of data tours (or with concrete example)
%         DFS for tree -> linear is hard to read

\section{Framework}
\label{section:methodology}
Supporting analytical reasoning for a dataset through semantic tours requires semantically linked data.
These links can be part of the dataset already, manually constructed, or automatically generated.
In either case, the elements of the dataset need to be interrelated to allow for the composition of a data tour.
In cases where the links are already part of the dataset, it can be challenging to extract them in practice when allowing for arbitrary input.
Hence, we focus on automatically generating the links instead.
% This linked data can either be existing relations as part of the data, manually constructed, or generated automatically through a formalism that can dissect and connect the dataset into interrelated units for the automatic composition of a data tour.
% In the former case, the parts of the dataset that reflect this interlinked data have to be specifically marked and be compatible, which makes automatic construction theoretically possible, but practically difficult when allowing arbitrary inputs, which is why we focus on the latter cases.
In the following, we formalize a data and semantic space that remains consistent for non-linear, semantic tours and linear tours as a simplified edge case of the former.
With the formalism, we address our first research question, \researchQuestion{1}.

\begin{comment}
\begin{itemize}
    \item mathematical formulation of data space and algorithms/processes (semantic space)
    \item transfer function requirements (scalable)
    \item theory in detail with constraints
    \item extraction
    \item processing
    \item representation
    \item UI relation finding
    \item Tasks?
\end{itemize}
\end{comment}

\begin{comment}
\begin{boxCaseStudy}{}
\ourToDo{We illustrate our methodology with a running case study from law that demonstrates the practical application of the data and semantic spaces.
In particular, we investigate a case of bodily harm in Germany, where a hunter shoots a mushroom picker~\cite{jusMundlichePrufungIm2025}.
Refer to \autoref{appendix:case-study} for the case description.}
\end{boxCaseStudy}
\end{comment}

% \begin{figure*}
%     \centering
    
%     % \missingfigure[figwidth=\linewidth]{A schematic figure of the data space and how the four operators interact with it.}
%     \includegraphics[width=\linewidth]{figures/Data Space (annotated).png}
    
%     \caption{The data space contains the data points of all \textcolor{blueDatasets}{data sets}. The four operators (1)~\textcolor{orangeSelection}{select} relevant data points~(\selectionOperator{}), (2)~\textcolor{purpleStaging}{stage} them in a scene~(\stagingOperator{}), (3)~\textcolor{greenSequencing}{determine the order} of these \textcolor{grayScenes}{scenes}~(\sequencingOperator{}), and (4)~\textcolor{yellowTransition}{transition} between them~(\transitionOperator{}).}
%     \label{figure:data-space}
% \end{figure*}
\subsection{Data Space}
\label{section:data-space}
We extend the mathematical description from \dataToursFramework{}~\cite{mehtaDataToursDataNarratives2017} and model tours in terms of four operators based on scenes of datasets~(refer to~\autoref{figure:data-space}).
For this, let ${\mathcal{D} = \{D_1, ..., D_n\}},\ n\in\mathbb{N}$ be the collection of \textcolor{blueDatasets}{\textbf{datasets}} that form the basis of a tour.
Each dataset $D_i$ consists of arbitrary data points, ${D_i = \{d_1,\ ...,\ d_m\}},\ m\in\mathbb{N}$.
Since tours can arbitrarily combine these data points over all data sets, we define ${\mathcal{D^\cup} = \bigcup_{D_i\in\mathcal{D}}D_i}$ and ${\mathcal{D^P} = \mathcal{P}(\mathcal{D^\cup})}$ as all possible combinations of data points.
% For example, data points could be high-dimensional vectors from the training of a machine learning model, time steps from a time series measured by a sensor, or papers in a scientific citation network.
Datasets do not have to be on the same granularity level; they can also denote groups or aggregates. 
In the legal domain, data points in different datasets can be single tokens, compound expressions, abstract legal concepts, specific paragraphs of a law, or even entire laws, which can all be connected and grouped.
% Another example could be time series in predictive maintenance and a corresponding list of anomaly types and their time range.

\begin{table}
    %\setcellgapes{.025 cm}
    \renewcommand{\arraystretch}{1.2}
    \footnotesize
    \vspace*{-1mm}
    
    \centering
    
    \caption{\textbf{The four operators that construct data tours in the data space based on \emph{DataTours}~\cite{mehtaDataToursDataNarratives2017}.}}
    \label{table:data-space-operators}
    
    \begin{tabular}{@{}l@{\hskip 2pt}|@{\hskip 2pt}l@{\hskip 1pt}l@{\hskip 3pt}|@{\hskip 3pt}l@{\hskip 3pt}p{2.6cm}@{}}
        \toprule
        \textbf{Name} & \textbf{OP} & & \textbf{Definition} & \textit{where} \\
        \midrule
        \textcolor{orangeSelection}{Selection} & 
        \selectionOperator{}$\, :$ & 
        $\mathcal{D}\to\mathcal{D^P}$ & 
        $O_\mathrm{sel}(\mathcal{D})\coloneq \{d_j,\ldots\}$ & {\scriptsize
        ${i,j\in\mathbb{N}},\; 1{\le}i{\le}j{\le}\vert\mathcal{D^\cup}\vert, \newline d_i\in\mathcal{D^\cup}$ }\\
        \textcolor{purpleStaging}{Staging} & 
        \stagingOperator{}$\, :$ & 
        $\mathcal{D^P}\to\mathcal{S}$ &
        $O_\mathrm{sta}(\mathcal{D^P})\coloneq S_i$ & {\scriptsize
        $i\in\mathbb{N},\ 1{\le}i{\le}k,\ S_i\in\mathcal{S}$ } \\
        \textcolor{greenSequencing}{Sequencing} & 
        \sequencingOperator{}$:$ & 
        $\mathcal{S}\to\mathbb{N}$ & 
        $O_\mathrm{seq}(\mathcal{S})\coloneq i$ & {\scriptsize
        $i\in\mathbb{N},\ 0{\le}i{\le}k-1$ }\\
        \textcolor{yellowTransition}{Transition} & 
        \transitionOperator{}$\, :$ & 
        $\mathcal{S}\to\mathcal{S}$ & 
        $O_\mathrm{tra}(S_i)\coloneq S_j$ & {\scriptsize
        ${i,j\in\mathbb{N}},\; 1{\le}i,j{\le}m,\newline O_\mathrm{seq}(S_j)=O_\mathrm{seq}(S_j)+1$} \\
        \bottomrule
    \end{tabular}
    \vspace*{-6mm}
\end{table}

\textcolor{grayScenes}{\textbf{Scenes}} provide concrete views on these datasets.
Let the set of scenes be defined as $\mathcal{S} = \{S_1, ..., S_k\},\ k\in\mathbb{N}$.
To construct tours, the \dataToursFramework{} framework defines \textbf{four operators}, which we slightly adapt.
For the operators and their mathematical definitions, refer to \autoref{table:data-space-operators}.
Each scene has a single \textcolor{orangeSelection}{\emph{selection operator}},~\selectionOperator{}, that \emph{filters} the \emph{data} to display in that scene.
To determine the display of that scene, it has a single \textcolor{purpleStaging}{\emph{staging operator}},~\stagingOperator{}, that \emph{configures} the \emph{view properties} of that scene, including the viewport, zoom, layout, and visual encodings, to display the result of the selection operator.
The \textcolor{greenSequencing}{\emph{sequencing operator}},~\sequencingOperator{}, \emph{determines} the \emph{order} of the scenes and the \textcolor{yellowTransition}{\emph{transition operator}},~\transitionOperator{}, seamlessly \emph{animates} between the data selections and view parameters of two subsequent scenes.
In the legal domain, a scene can render the text of a single paragraph or display the hierarchy of a law using a tree map, for instance.

\begin{comment}
\begin{boxCaseStudy}{(contd.)}
\ourToDo{We can use the four operators to filter for and display the relevant facts from the case's description.
Let us consider the words from the description that represent data points and constitute a data set.
The selection operator,~\selectionOperator{}, selects those data points that represent facts, such as \textquote{J}, \textquote{Jäger}, \textquote{V}, and \textquote{Pilze [...] sammeln} together with their context.
The staging operator,~\stagingOperator{}, creates a scene from selected data points by visually highlighting the facts in context.
For example, \textquote{J} and \textquote{Jäger} appear together in a scene as:
\begin{displayquote}
\entity{orange}{7}{Person}{J} ist passionierter \entity{green}{11}{Occupation}{Jäger} und begibt sich eines Morgens ganz früh auf die Pirsch in sein Revier.
\end{displayquote}
The sequencing operator,~\sequencingOperator{}, determines the order of the resulting scenes by the appearance of words in the case description. Hence, the above scene appears before the following:
\begin{displayquote}
J hat mit seinem Schuss den \entity{orange}{7}{Person}{V}, der bereits früh morgens aufgebrochen war, um \entity{green}{11}{Occupation}{Pilze zu sammeln}, getroffen.
\end{displayquote}
The transition operator,~\transitionOperator{}, animates from the first to the second scene by replacing the text and specifying the indices at which to display visual fact highlights.}
\end{boxCaseStudy}
\end{comment}

\subsection{From Linear to Semantic Space}
\label{section:linear-to-semantic-space}
\begin{figure}[!tb]
    \centering
    
    % \missingfigure[figwidth=\linewidth]{A schematic figure of the semantic space and how it augments the data space.}
    \includegraphics[width=\linewidth]{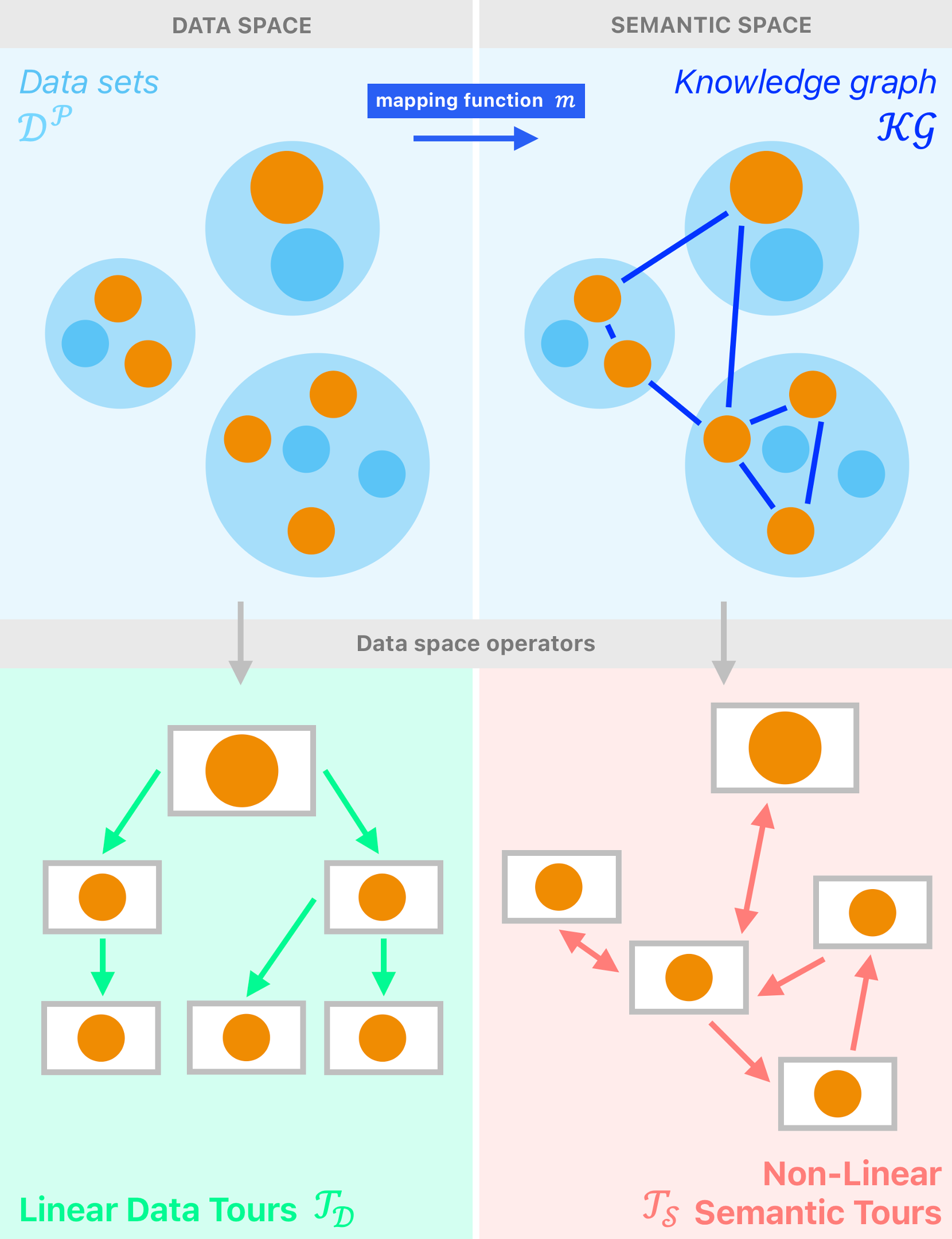}
    \caption{
        \textcolor{greenDataTours}{Data tours} traverse scenes of data points \textcolor{greenDataTours}{linearly} in the data space.
        Meanwhile, the semantic space \textcolor{blueKnowledgeGraph}{maps} data points to entities and connects them through semantic relationships, creating a \textcolor{blueKnowledgeGraph}{knowledge graph}.
        \textcolor{redSemanticTours}{Semantic tours} traverse the scenes of entities in a \textcolor{redSemanticTours}{non-linear} fashion.}
    \label{figure:semantic-space}
    \vspace*{-7mm}
\end{figure}
In \dataToursFramework{}~\cite{mehtaDataToursDataNarratives2017}, \citeauthor{mehtaDataToursDataNarratives2017} place \textcolor{grayScenes}{scenes} at a particular position in a single \textbf{tree} structure, reflecting their hierarchical level~(refer to~\autoref{figure:semantic-space}).
This tree can be defined as $T = (V_T, E_T)$, where $V_T = \mathcal{S}$ represents the scenes and $E_T$ represents the transitions between scenes.
The authors construct the path of tours through these scenes from a specific form of tree traversal, akin to a depth-first search, resulting in \textcolor{greenDataTours}{\textbf{linear data tours}}:
\begin{equation*}
    \mathcal{T}_D = \{(S_1, \dots, S_k) \mid S_j = O_\mathrm{tra}(S_i), i,j\le k, (S_i, S_j)\in E_T, S_i,S_j\in V_T\}
\end{equation*}
Here, $S_i$ and $S_j$ are two different scenes generated by the composition of the selection and staging operators, $O_\mathrm{sta}(O_\mathrm{sel}(\cdot))$.
It guides a viewer through a dataset in a defined order to present relevant insights.
Given the goal of \dataToursFramework{} to prompt data exploration, their choice of a tree is reasonable due to its advantages for providing an overview of and learning about a dataset~\cite{sarrafzadehKnowledgeGraphsHierarchies2016}.
For example, in the legal domain, a linear data tour could guide a viewer through scenes that portray the legal norms contained in a part of a law.

\par{\textbf{Non-Linearity of Sense-Making}} In our framework, \framework{}, we consider the semantic space in addition to the data space.
The semantic space caters toward sense-making, which partly overlaps with curiosity, comprehension, mental models, and situational awareness~\cite{kleinMakingSenseSensemaking2006}.
It focuses on understanding relationships between entities to make effective predictions about them in an iterative process~\cite{pirolli2005sensemaking} as modeled by the Data/Frame theory of sense-making~\cite{kleinMakingSenseSensemaking2006a}.
It assumes that frames represent an analysis perspective that defines relevant data, which also drives the change of that perspective, resulting in a closed loop~\cite{kleinMakingSenseSensemaking2006a}.
A frame can appear in various forms, such as a mind map, a diagram, or a story, all of which are meaningful in different analytical contexts.
The suitable representation for sense-making depends on the specific task~\cite{amadieuInteractionPriorKnowledge2010, sarrafzadehKnowledgeGraphsHierarchies2016}.
Since we address analytical reasoning in knowledge-centric domains, the natural way to express connections and relations is through a \textbf{graph}.
It is more effective in capturing the constituting entities and relationships, and providing answers for direct search than a tree~\cite{sarrafzadehKnowledgeGraphsHierarchies2016}, as relationships, context, and dependencies are typically not linear.
In particular, we use a \textcolor{blueKnowledgeGraph}{\textbf{knowledge graph}} $\mathcal{KG}$, which improves comprehension and reduces disorientation for people with significant amounts of domain knowledge~\cite{amadieuInteractionPriorKnowledge2010}.
Although one could interpret the semantic space as just another dataset subject to \dataToursFramework{}, we emphasize it as a first-class citizen in the data space, in particular due to its non-trivial linearization.
With the separate structure, we can leverage the advantages of KGs in the design of the corresponding UI.
Still, we consider entities of the semantic space to be elements of the data space, too.
This enables us to construct the semantic space through the operators of the data space and form connections easily.
While the path of tours in \dataToursFramework{} aligns with a tree traversal, \framework{} expresses \emph{relationships between scenes as sub-graphs of the KG}.
Hence, \framework{} can represent tours as graphs rather than a single linear path in a tree.
This allows for unconstrained data exploration during information foraging, where our framework is more expressive of analytical reasoning than \dataToursFramework{}.
That is, data hidden from the exploration hinders its interpretation, which negatively impacts analytical reasoning~\cite{kleinMakingSenseSensemaking2006}.
% Therefore, exploring data without constraints is closer to analytical reasoning than constrained data exploration in \dataToursFramework{}.

\par{\textbf{Mapping Data to Semantics}} We model entities and their relations within and between datasets of $\mathcal{D}$ as a knowledge graph, which enables arbitrary semantic relationships at different levels.
A \textbf{directed multigraph} is suitable to formalize this KG, because the multigraph allows for multiple edges between the same pair of nodes~\cite{bollobasModernGraphTheory1998}.
Formally, our multigraph is defined as ${\mathcal{KG} = \bigl(V_{\mathcal{KG}}}, E_{\mathcal{KG}}\bigr)$, where $V_{\mathcal{KG}}$ is the set of nodes and $E_{\mathcal{KG}}$ is the set of edges.
In this setting, $V_{\mathcal{KG}}$ represents entities in the KG, originating from data points in the data space, and $E_{\mathcal{KG}}$ refers to their semantic relationships.
% TODO: refine the hierachy!?
To construct a KG from the data space, one could use the points of the data space as nodes and construct the relationships as edges.
However, to allow for the creation of more complex relations, one might want to have additional nodes not present in the original data space and also enrich the data points with meta information.
This allows introducing new data points during construction, which implicitly creates an extension to the existing data space, or allows for entities from the user's implicit knowledge later during interaction with the UI, without modifying the data space.
Additionally, providing access to the unfiltered data space is desirable, since prior filters could bias the interpretation, which users may not be aware of.
Therefore, we construct the nodes of the KG, $V_\mathcal{KG}$, as independent units, which typically have a direct equivalence to a single data point in the data space~$\mathcal{D}$. Consequently, the type of node is implicitly defined by the type of underlying data point.
To connect these entities through edges, i.e., semantic relationships, we need a \textcolor{blueKnowledgeGraph}{\textbf{mapping function}}:
\begin{center}
    $m:(\mathcal{D}\times\mathcal{D})\to(\mathcal{R},\mathcal{M})$,
\end{center}
where $\mathcal{R} = \{R_1,\ ...,\ R_p\},\ p\in\mathbb{N}$ represents the types of semantic relationships and $\mathcal{M} = \{M_1,\ ...,\ M_q\},\ q\in\mathbb{N}$ represents metadata.
The mapping function relates two entities, associating them with a \emph{type of relationship} and relationship type-specific \emph{metadata} through an information function ${i(e) := M_e}$, where ${e\in E_\mathcal{KG}}$, and ${M_e\in\mathcal{M}}$.
For instance, in the legal domain, a knowledge graph can relate legal norms of a law to each other based on how they jointly define a legal concept, such as bodily harm.

\par{\textbf{Traversing Semantics}} While these relationships can be independent, some may form a hierarchy through \emph{induced higher-level relationships}.
These form due to the inherent inter-dependencies between individual datasets.
For example, if dataset $D_1\in\mathcal{D}$ consists of legal paragraphs that are related through $R_1$ by topics they cover, and another dataset $D_2\in\mathcal{D}$ combines the paragraphs from $D_1$ to laws, an induced relationship $R_1^\prime$ results between the elements, i.e., laws, in $D_2$, regarding the occurrence of topics.
Through both these induced relationships and their semantic meaning, the semantic relationships between entities can induce a conceptual hierarchy on entities.
Respectively, the corresponding data points in the data space can relate through the same hierarchy.
Navigating the hierarchy through the mapped relationships and reducing their complexity remains a major contribution of our VA design concept.
% We consider these constructs to be aggregates of data points and part of the data space $\mathcal{D}$.
% For this, we define a set of entity types $\mathcal{E} = \{E_1,\ ...,\ E_o\}\subset\mathcal{D},\ o\in\mathbb{N}$ and a set of semantic relationship types $\mathcal{R} = \{R_1,\ ...,\ R_p\}\subset\mathcal{D},\ p\in\mathbb{N}$.
% Additionally, we span a semantic hierarchy of levels $\mathcal{L} = \{1, ..., q\}\subset\mathcal{D},\ q\in\mathbb{N}$ across nodes that we leverage to design the user interface later on.
% Let $M$ be a set of five mapping functions $\mathcal{M} = \{m_\mathcal{D}, m_\mathcal{L}, m_\mathcal{E}, m_\mathcal{R}\}$~(refer to \autoref{table:mapping-functions}).
% The first, $m_\mathcal{D}$, assigns an ordering to the data points of a node.
% The second function, $m_\mathcal{L}$, assigns each node to a hierarchical level.
% The third function, $m_\mathcal{E}$, assigns each node to an entity type.
% The fourth function, $m_\mathcal{R}$, relates entities by assigning pairs of nodes to a type of semantic relationship.
% The last function, $m_\mathcal{D}^{-1}$, extracts the data points of a node.
However, the mapping function $m$ already contributes to reducing complexity by converting between implicit and explicit knowledge~\cite{wangDefiningApplyingKnowledge2009} and supports essential tasks of analytical reasoning, in particular the foraging loop, which is concerned with data exploration~\cite{pirolli2005sensemaking, kaixuAnalyticProvenanceSensemaking2015}.
The implicit knowledge may be known only to domain experts or may already exist in the data, albeit in an unstructured format.
% What tasks does the mere structure of the semantic space enable?
% For example, patterns of collaborations in scientific citation networks may only be revealed through data science, but are already part of the network as edges.
For example, in law, relationships between legal norms may not be explicitly expressed within them but may be mentioned in prose form in court rulings.
Extracting this implicit knowledge in practice through the mapping function can be achieved by pattern matching or, more effectively, a natural language-based approach, such as through an LLM.
Corresponding to the mapping function, we can define the KG's components, $V_{\mathcal{KG}}$ and $E_{\mathcal{KG}}$, as:
\begin{center}
    $V_{\mathcal{KG}} = \{d_i\ \vert\ d_i\in\mathcal{D}\}$ and $E_{\mathcal{KG}} = \bigl\{m(v, w)\ \vert\ v,w\in V_{\mathcal{KG}}\bigr\}$.
\end{center}
Together with the data space operators, we can construct tours through scenes as a graph based on a sub-graph of the KG, resulting in \textcolor{redSemanticTours}{\textbf{non-linear semantic tours}}:
\begin{align*}
    \mathcal{T}_S &= \bigl(V_{\mathcal{T}_S}, E_{\mathcal{T}_S}\bigr)\quad where \;\;  V_{\mathcal{T}_S} = \bigl\{O_\mathrm{sta}(v)\ \vert\ v\in V_\mathcal{KG}^\prime\bigr\} \qquad \textrm{and} \\
    E_{\mathcal{T}_S} &= \bigl\{(v, w)\ \vert\ w = O_\mathrm{tra}(v),\ v,w\in V_{\mathcal{T}_S},\ (v, w, \cdot)\in E_{\mathcal{KG}}\bigr\}
\end{align*}
with $V_\mathcal{KG}^\prime\subset V_\mathcal{KG}$, since semantic tours only consider sub-graphs of the KG.
Through this definition, we can also collapse non-linear semantic to linear tours if we source the edges without the KG:
\begin{center}
    $E_{\mathcal{T}_S} = \bigl\{(v, w)\ \vert\ w = O_\mathrm{tra}(v),\ v,w\in V_{\mathcal{T}_S}\bigr\}$,
\end{center}
thereby establishing \dataToursFramework{} as a simplified edge case of \framework{}, which is a specific, single linear path.
For example, in the legal domain, a semantic tour can leverage the legal norms linked by the KG to support drafting the legal argumentation.

%%%%%%%%%%%%%%%%%%%%%%%%%%%%%%%%%%%%%%%%%%%%%%%%%%%%%%%%%%%%%%%%%%%%%%%%%%%%%%%%%%%%%%%%%%%%%%%%

\section{Tasks and Design Rationales: A Scoping to Law}
\label{section:tasks-and-design-rationale}
In the following section, we discuss the design rationale behind our VA concept based on \framework{}.
We also discuss the requirements and tasks that our concept aims to address.
Due to the breadth of potential topics, we scope our studies on the law domain as one particular example of a knowledge-centric domain, intending to extend it in the future~(see \autoref{sec:future_work}).
% This then informs the construction of the UI and interaction concepts and the creation of tours in \autoref{section:user-interface}.

\begin{table*}[!tb]
    \centering
    \caption{A selection of analysis tasks in legal exploratory analysis based on the literature and informed by domain experts~(refer to~\autoref{section:tasks-and-design-rationale}).}
    \label{tab:tasks}
    \fontsize{7pt}{9pt}\selectfont
    \renewcommand{\arraystretch}{1.1}
    \begin{tabular}{@{}p{0.15cm}|r@{\hskip3pt}lp{6.15cm}p{5.6cm}@{}}
        \toprule
         & \textbf{\#} & \textbf{Task\textsuperscript{[source]}} & \textbf{Description} & \textbf{Exemplary Question} \\
        \midrule
        \multirow{2}{*}{\rotatebox[origin=c]{90}{\fontsize{5.3pt}{8pt}\selectfont Facts}}%
 & \taskFactExtraction{} & Facts extraction\textsuperscript{\cite{burtonThinkLawyerUsing2017}} & Identify legally relevant facts from narratives and judgments & \emph{Who is involved in the case?}  \\
 & \taskCounterfactuals{} & Counterfactuals\textsuperscript{\cite{lagnadoCausationLegalMoral2017}} & Hypothesize about outcomes under excluded and inverted facts & \emph{What would have happened if the injured party died?} \\
        \midrule
        \multirow{3}{*}{\rotatebox[origin=c]{90}{\fontsize{5.3pt}{8pt}\selectfont Norms}}%
 & \taskDoctrinalization{} & Doctrinalization\textsuperscript{\cite{ruethersRechtstheorieUndJuristische2022}} & Categorize a case into one or more legal fields & \emph{What legal fields does this case touch?} \\
 & \taskIssueIdentification{} & Issue identification\textsuperscript{\cite{burtonThinkLawyerUsing2017, ruethersRechtstheorieUndJuristische2022}} & Identify applicable statutes, case law,  commentary, etc. & \emph{Which legal norms from criminal law are relevant?} \\
 & \taskConflictResolution{} & Conflict resolution\textsuperscript{\cite{bundesministeriumfuerjustizHandbuchRechtsfoermlichkeit2024}} & Determine authority, hierarchy, binding force (lex specialis$\ldots$) & \emph{Which legal norm for bodily injury takes precedence?} \\
        \midrule
        \multirow{3}{*}{\rotatebox[origin=c]{90}{\fontsize{5.3pt}{8pt}\selectfont Subsumption}}%
& \taskNormInterpretation{} & Norm interpretation\textsuperscript{\cite{burtonThinkLawyerUsing2017, ruethersRechtstheorieUndJuristische2022}} & Interpret a legal norm using different methodology & \emph{How can we interpret this norm in the context of others?} \\
  & \taskArgumentConstruction{} & Argument construction\textsuperscript{\cite{bundesministeriumfuerjustizHandbuchRechtsfoermlichkeit2024}} & Draft complete arguments and counterarguments & \emph{How do the norms interplay in this case?} \\
% & \task{8} & \textcolor{gray}{Dialectical reasoning} & \textcolor{gray}{Weigh competing arguments and rebuttals systematically} & ... \\
  & \taskRuleConclusion{} & Rule-conclusion\textsuperscript{\cite{burtonThinkLawyerUsing2017}} & Derive the doctrinal outcome under the relevant legislation & \emph{What is the punishment for this crime?} \\
        \midrule
        \multirow{1}{*}{\rotatebox[origin=c]{90}{\fontsize{5.3pt}{8pt}\selectfont Meta \hspace*{-.8mm}}}%
% & \task{9} & \ourToDo{Comparative reasoning\textsuperscript{\cite{bundesministeriumfuerjustizHandbuchRechtsfoermlichkeit2024}}} & \ourToDo{Use comparative law/case to inform interpretation/precedence} & \ourToDo{\emph{Which line of argument did a previous court follow?}} \\
 & \taskAcademicCritique{} & Academic critique\textsuperscript{\cite{ruethersRechtstheorieUndJuristische2022}} & Evaluate doctrinal coherence, policy effects; propose reforms & \emph{Is the theory to interpret this norm missing?} \\
        \bottomrule
    \end{tabular}
\end{table*}

\par{\textbf{Legal Analysis}} Legal analysis follows a structured, iterative process involving several interconnected tasks.
To exemplify this process, we observe a German second state examination in law~\cite{briedenMundlichePrufungIm2025, jusMundlichePrufungIm2025}, which resembles practical bar examinations and covers a wide range of tasks that law experts perform during case analysis.
Although we source the list of tasks from the context of German law, they largely overlap with models from other jurisdictions, such as IRAC~\cite{burtonThinkLawyerUsing2017}.
Based on literature findings, we derive four categories of tasks~(refer to \autoref{tab:tasks}) that are relevant to case analysis.
\par{\textbf{Facts}}
The process begins when experts~\taskFactExtraction{}~identify legally relevant facts from the case~\cite{burtonThinkLawyerUsing2017}.
Consider our example where candidates note that a hunter shoots a mushroom picker with their rifle.
To determine relevance, experts~\taskCounterfactuals{}~also exclude or inverse facts to hypothesize about different legal outcomes~\cite{lagnadoCausationLegalMoral2017}.
Both tasks are essential to understanding causation in legal reasoning.
\par{\textbf{Norms}}
Experts then~\taskDoctrinalization{}~categorize the case into applicable legal areas~\cite{ruethersRechtstheorieUndJuristische2022}.
In our example, candidates associate the case with legislation found in criminal, civil, and gun laws.
These initial steps require our UI design to enable users to annotate or highlight facts in legal documents and provide visual associations between legal documents and legal areas.
Within said legal areas, experts~\taskIssueIdentification{}~identify applicable norms, court rulings, and commentary~\cite{burtonThinkLawyerUsing2017, ruethersRechtstheorieUndJuristische2022} while simultaneously~\taskConflictResolution{}~resolving conflicts between norms that apply concurrently~\cite{bundesministeriumfuerjustizHandbuchRechtsfoermlichkeit2024}.
The candidates in our example identify several criminal law norms involving bodily harm that could apply, but rule out some norms as being less specific than others.
These analytical tasks demand that our UI design enable users to look up legal artifacts, surface connections between them, and compare pathways among different chains of norms.
\par{\textbf{Subsumption}}
Legal experts must then~\taskNormInterpretation{}~interpret the remaining norms using different methodologies~\cite{burtonThinkLawyerUsing2017, ruethersRechtstheorieUndJuristische2022} and~\taskArgumentConstruction{}~argue how this interpretation applies to the specific case~\cite{bundesministeriumfuerjustizHandbuchRechtsfoermlichkeit2024}.
The methods can be \interpretationGrammatical{}~grammatical, \interpretationTeleological{}~teleological, \interpretationSystematic{}~systematic, or \interpretationHistorical{}~historical interpretation~\cite{ruethersRechtstheorieUndJuristische2022}; although others exist as well.
The candidates demonstrate this by using systematic reasoning to relate case facts to the requirements for applying their selected norm.
Finally, experts~\taskRuleConclusion{}~conclude their argumentation under the selected norms to rule the case~\cite{burtonThinkLawyerUsing2017}.
Reasoning and argumentation require our UI design to enable users to draw connections between annotations or highlights and legal artifacts, and to present their reasoning as an overview to others.
\par{\textbf{Meta}}
In some instances, legal experts also~\taskAcademicCritique{}~critique statutes, theories, and policies within the context of a concrete case, though this was not relevant to our example~\cite{ruethersRechtstheorieUndJuristische2022}.
Together, these tasks highlight the requirements that a system supporting legal analytical reasoning should satisfy.

%%%%%%%%%%%%%%%%%%%%%%%%%%%%%%%%%%%%%%%%%%%%%%%%%%%%%%%%%%%%%%%%%%%%%%%%%%%%%%%%%%%%%%%%%%%%%%%%

% Examples: Kann im Sinne des § 61 PBefG auch dem Kraftfahrzeugführer ein Bußgeldverfahren eröffnet werden? → BayObLG 29.06.2000 NZV 2000, 424: „Normadressat des § 61 I Nr. 1 i.V.m. § 2 I PBefG ist der Unternehmer. Andere Personen, insbesondere der nicht mit dem Unternehmen identische Kraftfahrzeugführer, können sich nur gemäß § 14 I OWiG an der Ordnungswidrigkeit beteiligen.“ (Mögliche Query: § 61 PBefG Bußgeldverfahren (Fahrer | Fahrzeugführer | Kraftfahrzeugführer))
% Examples: https://pubmedkg.github.io/ (https://www.nature.com/articles/s41597-025-05343-8)
% Examples: Wind Turbines in Denmark (https://diglib.eg.org/items/a6b851bb-49f3-4108-b065-d35500785d70)

% !!!!!
% InsideInsights has good literature on using hierarchies vs graph structures for different types of users! Use them! They call it "Dynamic Insight Hierarchies"
% !!!!!

\section{Visual Analytics Design Concept for \framework{} in Law}
\label{section:user-interface}

\begin{figure*}[!bt]
    \centering
    
    \includegraphics[width=\linewidth]{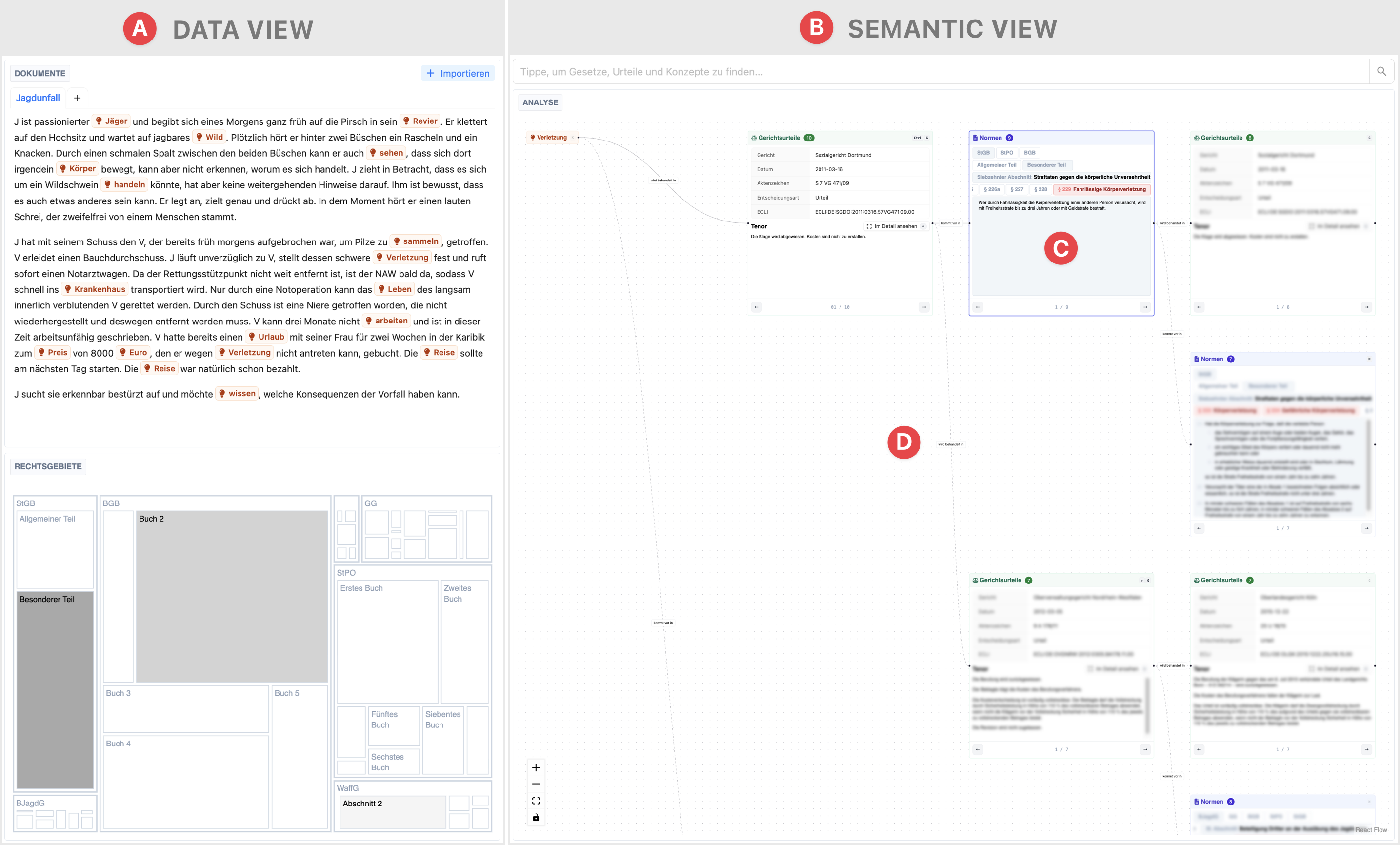}
    
    \caption{A concept draft of the user interface~(UI) of a Visual Analytics application using \framework{}. The UI consists of a data view~(A) and a semantic view~(B). The semantic view acts as a lens that users interactively move over the underlying knowledge graph, representing the iterative process subject to analytical reasoning. The semantic view represents the entities~(C) and relationships~(D) of the semantic tour.}
    \label{figure:design-draft}
    \vspace*{-5mm}
\end{figure*}

We \emph{conceptualize} a VA design for a system supporting legal analytical reasoning to address \researchQuestion{2}, based on the identified tasks~(refer to~\autoref{section:tasks-and-design-rationale}) and our framework~(refer to~\autoref{section:methodology}).
We refer to the German second state examination from \autoref{section:tasks-and-design-rationale} to exemplify parts of the workflow that our design supports.
The UI of the concept consists of two main elements: a data view on the left~(\autoref{figure:design-draft}, A) and a semantic view on the right~(\autoref{figure:design-draft}, B).

\subsection{Data View}
\begin{comment}
\begin{figure}[!tbh]
    \centering
    
    \missingfigure[figwidth=\linewidth]{The data view showing an overview of case and law, and showing the details of a visual highlight.}
    
    \caption{\ourToDo{The data view depicts the data space~(refer to \autoref{section:data-space}). Specifically, the data view surfaces the data points that are currently relevant to the semantic view. In this example, (A) the top part of the data view displays the text of a legal case, whereas (B) the bottom part shows the legal norms of a law.}}
    \label{figure:data-view}
\end{figure}
\end{comment}
The data view is the entry point to the legal analytical reasoning process.
% The data view presents data points of entities at the current position and their vicinity in the KG.
It displays legal text documents and their structure, showing the text itself~(\taskNormInterpretation{}/\interpretationGrammatical{}) and a tree map of the corresponding formal legal structure~(\taskNormInterpretation{}/\interpretationSystematic{}).
The text view automatically highlights relevant entities, enabling users to extract facts from the data~(\taskFactExtraction{}, \taskCounterfactuals{}).
Since entities may be missing from the KG, users can introduce new entities not yet part of the knowledge graph from existing data points.
For example, users can highlight a sentence in the legal text document that implies a relationship to other parts, thereby introducing a new entity~(\taskIssueIdentification{}, \taskNormInterpretation{}).
The tree map highlights the legal fields relevant to the legal document in the text view, providing an overview~(\taskDoctrinalization{}).
% Both approaches enable users to extract facts from the data~(\taskFactExtraction{}).
Through these highlights, the data view links to the semantic view.
To start the reasoning process, the user selects an entity of interest such as bodily harm~(\emph{Verletzung}) in \autoref{figure:design-draft}.
This selection guides the user to the semantic view.
% These highlights encode legal fields~(\taskDoctrinalization{}) and function as deep links that re-center the semantic view on the corresponding entity when clicked.

\subsection{Semantic View}

With an entity selected from the data view, the semantic view displays a node-link diagram of related entities.
These represent the neighborhood of the selected entity in the KG.
In our example, these entities refer to legal concepts, legal norms, and court rulings.
\par{\textbf{Visual Representation}
The related entities appear using data-specific representations.
For example, legal norms appear in an Icicle Plot that renders the corresponding law's structure~(\taskNormInterpretation{}/\interpretationSystematic{}) with the legal norm text at the bottom~(\taskNormInterpretation{}/\interpretationGrammatical{}).
Edges connect these representations as lines that encode semantic relationships~(\taskNormInterpretation{}/\interpretationSystematic{}) and their associated metadata.
The node-link diagram follows a layered layout where each row represents a chain of connected entities and columns represent steps in the analytical reasoning process.
Since nodes in a semantic tour may have many neighbors, our concept aggregates their representations to reduce visual clutter~\cite{shiScalableNetworkTraffic2013}.
For example, several legal norms connected to a common node appear in a joint hierarchy with laws at the top of the before-described Icicle Plot.
% Cells aggregating multiple neighbors with the same relationship type display a count badge; selecting it opens a disambiguation context.
\par{\textbf{Navigation}}
Starting from the selected entity, the current step of the reasoning process is indicated with an outline around its node.
In our example~(refer to~\autoref{figure:design-draft}), the third step of the reasoning process is outlined in the middle of the semantic view.
The navigation to the next step of the tour occurs through keyboard shortcuts and clicks on entities~(\autoref{figure:design-draft}, C) or relationships~(\autoref{figure:design-draft}, D)~(\taskNormInterpretation{}/\interpretationSystematic{}, \taskArgumentConstruction{}).
% Users can also change the relationship type represented by a row after selecting from a list.
% Scrolling enables movement along the \textbf{hierarchy} induced by relationships~(\taskDoctrinalization{}, \taskConflictResolution{}).
% For instance, when two laws are implicitly related through legal norms that reference each other, users can click the implicit relationship to focus on those norms, then scroll upwards to return to the law-level relationship.
% The \textbf{semantic breadcrumbs}~(\autoref{figure:design-draft}, F) provide an alternative navigation method for the same hierarchy~(\taskDoctrinalization{}, \taskConflictResolution{}) but provide context not available through scrolling alone.
If a relationship is missing from the semantic view, users can edit the relationships between entities to integrate their \textbf{tacit knowledge}~(\taskIssueIdentification{}, \taskNormInterpretation{}) similar to the data view.
Since these relationships are inherently more complex in graph-based tours than in linear ones, our concept employs a semantic lens to guide navigation over the tours.
The lens focuses on the current reasoning step, while blurring distant entities that have not been visited and are currently not relevant~\cite{kosaraSemanticDepthField2001}.
\autoref{figure:design-draft} illustrates this design with unveiled entities at the top and others still being blurred.
Moving along the relationships unveils those entities, creating reasoning paths within a semantic tour.
In turn, these reasoning paths create \textbf{analytical provenance}, allowing the analyst to move between recently visited entities~(\taskConflictResolution{}, \taskNormInterpretation{}/\interpretationSystematic{}).
\begin{comment}
\par{\textbf{Analytical Provenance}}
The semantic view includes a mini map in the lower right corner~(\autoref{figure:design-draft}, D) that provides context on analytical provenance.
By default, it presents an overview of semantic paths taken~(\taskArgumentConstruction{}, \taskRuleConclusion{}).
Clicking the mini map enlarges it to a full-screen modal where entities become deep links similar to data view highlights. 
A search bar above the node-link diagram~(\autoref{figure:design-draft}, E) enables semantic search of KG entities~(\taskFactExtraction{}, \taskIssueIdentification{}).
Selecting search results moves focus to the chosen entity.
The search bar also allows users to select entities for tentative relationships in the node-link diagram~(\autoref{figure:design-draft}, TODO)~(\taskIssueIdentification{}, \taskNormInterpretation{}).
\end{comment}

\subsection{Semantic Tours}
\begin{comment}
\begin{figure}[!tbh]
    \centering

    \missingfigure[figwidth=\linewidth]{A figure showing a user revisiting analytical provenance by replaying a tour from the mini map.}
    
    \caption{\ourToDo{\framework{} enables branches in analytical reasoning and supports detours to distant entities. Most importantly, the UI enables the user to revisit analytical provenance by replaying a tour from the mini map.}}
    \label{figure:branches-and-detours}
\end{figure}
\end{comment}
A non-linear, semantic tour, akin to the Data/Frame theory~\cite{kleinMakingSenseSensemaking2006a}, begins with a minimal frame that provides focus within the vast KG.
\par{\textbf{Initialization}}
We construct this initial frame in our concept \textbf{based on the user's context}, sourced from either a provided document or an entity selected through direct search.
When users select a specific entity, the tour begins with the semantic view centered on that entity.
In the provided documents, we match entities and their relationships from the document with the KG, like bodily harm in \autoref{figure:design-draft}.
This matching process can produce distinct entity groups, leading us to generate different semantic tours for exploration.
The analytical reasoning process unfolds as the user \textbf{navigates along relationships} between entities, creating a semantic path that schematizes the data~(\taskFactExtraction{}, \taskArgumentConstruction{}).
% Initial navigation proceeds automatically based on semantic relationship relevance, though users can intervene at any time to continue manually.
\par{\textbf{Branches and Detours}}
As reasoning progresses, semantic paths expand through branching and detouring at various entities of the tour.
\textbf{Branching} occurs when users consider entities adjacent to a semantic path.
In our example of bodily harm, the lawyer could consider a related court ruling that covers a similar case or opposing opinion~(\taskCounterfactuals{}).
\textbf{Detouring} happens when users perform direct searches for entities not part of or reachable from the current semantic tour~(\taskCounterfactuals{}).
In \autoref{figure:design-draft}, this would be the case when selecting another entity other than bodily harm, such as vacation~(\emph{Urlaub}), to consider indirect civil damages incurred by the bodily harm during the analysis.
These branches and detours allow us to define semantic tours in terms of \textbf{semantic paths}~--~groups of reasoning steps that divide tours into comprehensible components.
However, semantic paths remain non-linear due to elements such as replays.
\par{\textbf{Replay}}
The process of searching for and schematizing data points involves hypothesizing about data interpretation.
This includes \textbf{revisiting analytical provenance} to gain an overview of the semantic tour and exploring semantic paths.
Legal experts use replays to construct their argument for a specific case, as in our example~(\taskArgumentConstruction{}).
From the overview, analysts can interactively replay tours to question and reframe their evolving frame.
That is, an analyst moves to previously visited legal norms, interprets them systematically~(\taskNormInterpretation{}/\interpretationSystematic{}), and revises their arguments~(\taskArgumentConstruction{}).
When encountering dead ends, analysts can hypothesize about relationships to other entities by forging new connections~(\taskIssueIdentification{}).
As the reasoning process concludes, legal experts utilize the node-link diagram of the semantic view to present their analytical reasoning~(\taskRuleConclusion{}), leveraging the complete semantic tour as a foundation for communicating their findings.

\section{Preliminary Evaluation}
\label{sec:user_study}
% TODO: a note to the reviewers about the requirements of our research environment regarding studies involving human subjects is necessary! (refer to https://iui.hosting.acm.org/2026/call-for-papers)
% What practical impact do experts see in our application?

% Lucas: think from the research question -> what do you want to show? -> decide on the appropriate study task(s)
%        for IUI (and other more rigorous conferences), having p-values, leveling for effects, and so on is definitely necessary
%        if the story is to improve upon legal analytical reasoning, show that with the study: compare the existing approach (e.g., law on paper + beck-online) with your approach
%        as discussed previously, a comparison between data tours and semantic tours is not really fair -> a study comparing the two would not really make sense -> the storyline needs to change as well
%        time to deadline of IUI is tight, but could also aim for PacificVis Conference Paper Track (deadline at the beginning of November) or EuroVis Full Paper Track (deadline at the end of November)
%        in general, make sure to be content with the paper storyline

We conduct a preliminary expert user study with $n = 6$ participants to evaluate \framework{}~(refer to~\autoref{section:methodology}), the tasks~(refer to~\autoref{section:tasks-and-design-rationale}), and our conceptualized design~(refer to~\autoref{section:user-interface}) for a VA application scoped to law. A specific focus is set on the support that our framework can provide for analytical reasoning in the legal domain.

\subsection{Participants and Methodology}
% The pilot study \ourToDo{(see section )} with experts from jurisprudence~\cite{todo} identified beck-online~\cite{verlagc.h.beckohgBeckonline} as the primary means for legal research.
% Besides beck-online, the experts also reported that paper-based printouts of legal documents play a role in legal research.
% For the legal domain, these experts are highly specialized in specific areas.
% While such experts have diverse backgrounds, their predisposition toward legal areas is not suitable for our user study.
% Hence, we work with students of law who have not specialized in a legal area but are knowledgeable generalists.
% To cover a broad range of legal specializations, 
% These experts represent professionals from legal practice, including courts and law firms.

We interviewed six legal experts~(\expert{1}-\expert{6}), with three of them from the German legal system~(\expert{1}-\expert{3}) and the remainder from the Dutch legal system~(\expert{4}-\expert{6}).
Four participants identify as female, while the other two~(\expert{2} and \expert{3}) identify as male.
% 30 + 27 + 31 + 45 + 28 + ?
The average age of the participants, who chose to provide their age, was 32.2~years~(SD~=~7.328, $n = 5$).
On the German side, \expert{1} is a tax consultant and lecturer with a Master's degree in law~(LL.M.), 1.5 years of experience as a lecturer and 4 years of professional experience.
\expert{2} is a public official with a degree in public management, an administrative and public law background, and nearly 2 years of professional experience.
\expert{3} is a lawyer who passed the second German state examination, specializing in civil and public law, and has 1.5 years of professional experience.
On the Dutch side, \expert{4} is a lawyer with an LL.M. in criminal and governance law, and more than 20 years of professional experience.
\expert{5} is an administrative lawyer with an LL.M., a focus on administrative and public law, and 1 year of professional experience.
\expert{6} is a civil lawyer with an LL.M., a background in public law, and 14 years of professional experience.

\par{\textbf{Methodology}}
We conducted the user study in a controlled environment, with sessions lasting between 45 and 60 minutes, and no compensation was provided for participation.
% After obtaining consent, we recorded the audio of the remaining study procedure and participants' interactions with our implementation of \framework{}.
Each session began with the collection of demographic data and an interview on participants' experiences with traditional legal analytical reasoning. % and Visual Analytics.
% Afterward, we introduced our framework implementation through a pre-recorded video that explained the developed navigation and interaction paradigms, specifically mapping these paradigms to the tasks and requirements of legal analytical reasoning~(refer to \autoref{section:tasks-and-design-rationale}).
% Following the video, participants engaged in a 10-minute training phase where they could freely explore the implementation independent of the case at hand while we answered questions about the navigation and interaction paradigms.
We then introduced the tasks that ground our VA concept (refer to~\autoref{section:tasks-and-design-rationale}) and asked participants to relate them to their legal experience.
Next, we asked them to envision an application that supports these tasks in their legal analysis.
Afterwards, we presented the data and semantic views from our VA design (refer to~\autoref{section:user-interface}) that instantiates our model (see \autoref{section:methodology}).
For graph-based tours, we also presented a linear alternative to the semantic view; for the data view, several alternatives were shown (refer to the supplemental material).
Finally, participants assessed the potential real-world impact of the design.
% We emphasized that the task had no deterministic solution and that we would not judge their legal reasoning.
% We instructed the participants to perform the task for 30 minutes while thinking aloud, with particular focus on sharing any insights they gained throughout the process.
% To orient the participants throughout the task, we used the guiding questions from above.
% The session concluded with participants completing a SUS questionnaire, followed by a qualitative interview~(refer to \autoref{appendix:qualitative-interview}) to gather additional insights about their experience.

\subsection{Results}

Although our participants come from different legislative systems and legal fields, they share common methods of legal analytical reasoning.
% EXPERIENCE WITH TRADITIONAL LEGAL ANALYTICAL REASONING METHODS (+ CHALLENGES)
\par{\textbf{Tasks}}
All experts agree with our categories, selection, and order of tasks~(refer to~\autoref{section:tasks-and-design-rationale}) and find them relevant.
Still, there are individual divergences.
For example, \expert{5} starts by \quote{clearly defining and narrowing down the legal issue} before diving into fact extraction.
\expert{1} and \expert{3} emphasize that the importance of different tasks can be case-specific, e.g., unknown legal issues require a deeper comprehension of legal materials than well-known constellations.
\expert{4} also notes that some cases require the application of technical knowledge or knowledge from third-party domains, which can pose a challenge.
% CHALLENGES FACED
The experts also reported facing other obstacles.
For example, \expert{2} finds extracting relevant facts from court rulings hard, especially if their verdict is long-winded.
\expert{3} struggles in getting to know the established keywords for unknown legal sub-areas, and \expert{5} suffers from constant switching between different legal sources.
% IMAGINE A SUPPORTING APPLICATION
\par{\textbf{Ideal Application}}
These hardships depend on the working media, which differ across participants.
\expert{1} and \expert{4} prefer paper sources, \expert{3}, \expert{6} and \expert{5} work with domain-specific software, and \expert{2}, mixes analog and digital tools based on the case.
To overcome these obstacles, our participants envision an environment that \textbf{automatically extracts facts}~(\expert{3}), enables \textbf{rich annotations}~(\expert{2}), supports \textbf{continuity}~(\expert{1}), and \textbf{reduces context switches}~(\expert{5}).
\expert{3} stresses the burden of extracting relevant facts and calls for automation, accompanied by preliminary recommendations on relevant legal norms, commentary, and court rulings~(\expert{3} and \expert{5}).
\expert{4} stresses the importance of guarding against bias by surfacing opposing perspectives alongside prevailing views.
\expert{5} imagines that a \quote{visual map of how legal terms, norms, and cases connect would be very helpful.}
Having such a visualization, \expert{2} wishes for word-level annotation capabilities that link individual words to sections of norms, court rulings, and legal commentary, and allow them to externalize their tacit knowledge as comments.
Like \expert{1} and \expert{5}, they also want to bookmark frequently referenced legal norms and other relevant legal documents.
% CONCEPTUALIZED VISUAL ANALYTICS DESIGN
\par{\textbf{VA Design Concept}}
Introduced to our VA design concept~(refer to~\autoref{section:user-interface}), the participants had different preferences for data views.
% DATA VIEW
% VARIANT 1 (TREEMAP)
While some take value from the tree map in our concept~(variant~1)~(\expert{4}-\expert{6}), participants considered the visualization to be overwhelming~(\expert{2} and \expert{3}) and too abstract lacking the display of norm labels~(\expert{1}, \expert{2}, and \expert{6}).
% VARIANT 2 (SEMANTIC MAP)
The German domain experts~(\expert{1}-\expert{3}) agreed that the semantic map~(variant~2) is too overwhelming for them due to the crowded circles.
The Dutch legal experts~(\expert{1}-\expert{3}) also struggled with the noise in the visualization but appreciated it for its overview and could anticipate a use case if the semantic map would indicate the current position during navigation~(\expert{6}).
% VARIANT 3 (WORD CLOUD (OURS))
The participants appreciated the simplicity of the word cloud~(variant~3) as it \quote{gives a better intuition overall and more insights into relevant concepts}~(\expert{4}).
\expert{3} imagined a symbiosis between the tree map and the word cloud, with the former providing an overview and the latter giving details on demand.
Though they would like the legal norms to be annotated in the word cloud.
\expert{4} could also imagine combining the structure exhibited by the tree map with the word cloud.
Both \expert{4} and \expert{6} could imagine using the same visualization technique for documents rather than concepts.
% VARIANT 4 (WORD CLOUD (BECK-ONLINE))
All participants preferred variant~3 with its color coding and visual spacing over variant~4, finding it less cluttered.
% SEMANTIC VIEW
\par{\textbf{Semantic View}}
Comparing the two semantic views, participants expressed a preference for the non-linear graph view~(variant 1) over the linear list~(variant 2).
The experts found the latter to be cluttered due to a lack of structure, which they appreciated in the former because of its separation of documents.
\expert{6} finds it \quote{easy to see everything in an intuitive way} with \quote{all of the documents [being] well ordered.}
\expert{2} appreciates the widgets for documents as they \quote{already provide a preview of [a court ruling's] tenor.}
Despite the separation of documents, the graph view also links these, which \expert{5} finds \quote{more intuitive than scrolling through long documents or clicking endlessly through different databases.}
\expert{1} agrees, finding guidance and orientation in the connections between documents, which they do not experience with the list.
\expert{3} concurs, finding value in the connections as they can indicate potential chains of legal norms.
Generally, \expert{4} remarks that \quote{despite the linear nature of [...] law[, i]t might be necessary to take side-paths, so while it might start out as a tree, it should be seen more as a graph.}
Still, \expert{2} and \expert{3} warn that the visualization must ensure that documents are ordered by relevance.
Otherwise, as they would manually need to step through the documents, they would find themselves in the same situation as with the list.
\expert{5} expects the view to become overwhelming easily, and \expert{4} cautions that the application confines the user to its selection of documents without considering others.
Hence, the experts wish for more visual orientation and guidance~(\expert{3}-\expert{6}).
For example, \expert{3} suggests semantic breadcrumbs, and \expert{5} would like to see the current path highlighted.
% REAL-WORLD IMPACT
\par{\textbf{Real-World Impact}}
Regardless, \expert{5} finds that \quote{the real-world impact could actually be quite big, especially for people who regularly switch between different laws, domains and case law}, referring to our VA design concept and finding it \quote{genuinely innovative.}
\expert{4} agrees, seeing that legal research \quote{would be more holistic and consider a broader range of perspectives.}
\expert{3} joins, foreseeing that they \quote{could quickly familiarize [themselves] with a case}, especially since they work with different legal areas.
While \expert{6} does not disagree, they warn that \quote{to get the most out of these applications, you want to create as many connections as possible with [...] internal documentation systems, and in practice this turns out to be a significant obstacle.}

\begin{comment}
\begin{figure}[!tbh]
    \centering

    % https://bib.dbvis.de/uploadedFiles/Joos_eyetracking_ICMR_2024.pdf
    \includegraphics[width=.4\linewidth]{figures/Heat maps.png}
    
    \caption{Recording the amount of interaction with different parts of our implementation's UI, as well as with the different entities and relationships from the study's KG, shows interesting patterns of participants engaging with analytical reasoning.}
    \label{figure:study-heat-map}
\end{figure}
\end{comment}

%%%%%%%%%%%%%%%%%%%%%%%%%%%%%%%%%%%%%%%%%%%%%%%%%%%%%%%%%%%%%%%%%%%%%%%%%%%%%%%%%%%%%%%%%%%%%%%%

\section{Discussion}
Based on our preliminary evaluation, we elicit findings and discuss lessons learned for \framework{}.

\subsection{Findings and Lessons Learned} % Lessons Learned = Design Implications (https://arxiv.org/pdf/2407.12192)
Overall, the feedback from the evaluation of \framework{}, the tasks, and our conceptualized design for VA was encouraging.
% MODEL
\par{\textbf{Model}}
Our graph-based framework and the resulting VA design concept appear to reflect the analytical reasoning workflow that domain experts typically employ.
However, through the design concept, we only implicitly tested the model.
An implementation in the legal domain would be beneficial for stronger evidence.
In particular, to verify the practicality of the \textcolor{blueKnowledgeGraph}{mapping function}.
% TASKS
\par{\textbf{Tasks}}
The tasks accurately reflect those encountered by domain experts in legal practice.
Despite nuanced differences, the tasks and their categories apply to a wide range of legal fields across the Dutch and German legal systems.
Our selection will be useful for researchers to design VA applications that support legal practice.
% DATA VIEW
\par{\textbf{Data View}}
Among the various visualization techniques presented in the evaluation, a preference emerged for the word cloud.
Some participants also brought up the idea of linking it to the tree map as a combination that provides an overview and details on demand.
However, one of the visualizations should include labels for the legal norms as the experts deem them relevant.
% SEMANTIC VIEW
\par{\textbf{Semantic View}}
Our evaluation unveils that legal experts find value in graph-based tours of legal documents.
In particular, they appreciate the aggregated graph nodes that simplify the inherent complexity of such tours, showing different types of legal documents.
The experts also enjoy the concept of a semantic lens that displays a subset of the KG.
To experience the navigation through that lens, an implementation of the design is necessary.
% EVALUATION
\par{\textbf{Evaluation}}
While our preliminary evaluation confirms insights from a previous user study that evaluated a three-phase workflow for VA in law~\cite{furstChallengesOpportunitiesVisual2025}, we recognize the need for a quantitative user study that has legal experts conducting the analysis tasks~(refer to~\autoref{section:tasks-and-design-rationale}) in a comparative setting using our VA design concept and traditional methods like a paper copy of law.
Such a setting would provide more robust evidence of the usefulness of the design variants for data and semantic views in various legal tasks.
The evaluation would also benefit from a fully implemented design concept, as this would put more focus on the interaction design, which is not immediately apparent from the concept alone.

\subsection{Limitations and Future Work}
\label{sec:future_work}
Beyond the evaluation, our work on \framework{} opens several research directions that span implementation, agentic tours, dynamic graphs, and domain application.

\par{\textbf{Prototype}}
While we have conceptualized a VA design for \framework{} in law, we have not provided an implemented prototype.
Creating a prototype would necessitate finding a suitable \textcolor{blueKnowledgeGraph}{mapping function} for a legal KG, which can be a challenging task.
Still, a test-case-based evaluation would surface insights about the practicality of our concept and facilitate more detailed feedback from legal experts.
The evaluation of the design concept has already yielded promising feedback, suggesting its potential usefulness in legal practice.
Additionally, an implementation would enable a thorough evaluation of the interaction design in our concept.

\par{\textbf{Agentic Tours}}
We deliberately excluded Artificial Intelligence~(AI) agents from our framework to focus on the contrast between linear and non-linear tours.
A VA application based on our framework should incorporate the prevailing opinion within a spectrum, as noted by legal experts in the preliminary evaluation.
In a mixed-initiative system~\cite{horvitzPrinciplesMixedInitiativeUser1999}, AI agents could suggest semantic pathways based on user context, proposing alternative exploration routes at branching points and highlighting potentially relevant but unexplored relationships~\cite{kimLegisFlowEnhancingKorean2025, thenmozhiHierarchicalKnowledgeGraph2025}.
Designing a VA application that accommodates such agents is an open research question.
% A central research in this idea revolves around the balance and timing of automation with human autonomy.
Previous work suggests that domain experts need to experience exploration themselves, as external interpretations can hinder analytical reasoning~\cite{kleinMakingSenseSensemaking2006}.
Therefore, future work should investigate how AI agents can enhance rather than replace human reasoning, particularly in determining when and how algorithmic suggestions prove beneficial rather than detrimental to analytical thinking.

\begin{comment}
\par{\textbf{Human-Human Collaboration}}
Semantic tours could transform how experts collaborate on complex analytical problems.
The provenance captured in our mini map creates natural starting points for knowledge sharing and collaborative reasoning. 
Research opportunities include asynchronous collaboration, where experts build upon others' analytical pathways; synchronous collaborative exploration of semantic spaces; and integration with existing professional tools to support knowledge preservation.
A key challenge involves maintaining the benefits of individual reasoning while enabling effective collaboration. 
Research questions arise around how to represent multiple expert perspectives on semantic relationships, resolve conflicting interpretations, and design interfaces that support both individual and collaborative reasoning processes.
Additionally, investigating how semantic tours could preserve and transfer tacit knowledge from experienced practitioners to novices represents an opportunity for organizational learning.
\end{comment}

\par{\textbf{Dynamic Graphs}}
Currently, semantic tours operate on static knowledge graphs, but many domains involve evolving knowledge structures. 
Introducing a temporal dimension would enable reasoning about how concepts, relationships, and interpretations change over time.
In law, this could support analysis of legal doctrine evolution through court decisions and legislative changes.
In digital humanities, temporal semantic tours could trace the development of historical concepts and their relationships across different periods.
Technically, this requires extending our knowledge graph model with temporal metadata for entities and relationships, along with an interface design that supports temporal navigation.

\par{\textbf{Domain Application}}
While we have demonstrated \framework{} in the legal domain, the underlying framework~(refer to~\autoref{section:methodology}) is domain-agnostic.
This generality suggests applications across knowledge-centric domains where experts reason about complex relationships between entities.
Promising domains include software security analysis, where experts trace data flow to identify vulnerabilities; predictive maintenance, where anomaly patterns relate to equipment failure; and medical diagnosis, where symptoms, conditions, and treatments form complex semantic networks.
However, each domain requires an appropriate mapping function to construct the semantic space.
Additionally, domain-specific interaction paradigms that differ from our legal interface design may be necessary.
Hence, a critical research question arises regarding the characteristics that make a domain suitable for semantic tours.
Understanding the cognitive and structural requirements could guide systematic expansion to new fields and inform design principles for domain-specific implementations.

% Research opportunities include optimizing cognitive load through improved information architecture, personalizing interfaces to accommodate different cognitive styles and expertise levels, and understanding the long-term effects of our framework usage on expert reasoning patterns.
% Our current node-link diagram approach may not be optimal for all relationship types or user preferences.
% This holds for \framework{} as a domain-agnostic framework for analytical reasoning, but also for the particular application domain of law.
% In the latter, previous works have predominantly explored node-link diagrams and text~\cite{todo}.
% Future work should explore alternative representations and adaptive visualizations that respond to evolving user context and task requirements.

%%%%%%%%%%%%%%%%%%%%%%%%%%%%%%%%%%%%%%%%%%%%%%%%%%%%%%%%%%%%%%%%%%%%%%%%%%%%%%%%%%%%%%%%%%%%%%%%

\section{Conclusion}

We presented \framework{}, a conceptual model that extends data tours, grounded in a semantic abstraction of the data.
We formalized both the data space and the semantic space, connecting them through selection, staging, sequencing, and transition operators.
Through a user-definable mapping function, we construct a knowledge graph~(KG) from the data space that relates entities via semantic relationships.
In the KG, we establish semantic tours as sub-graphs that support the dynamic, non-linear nature of analytical reasoning compared to static, linear data tours.
This establishes data tours as a simplified edge case of semantic tours.
Building on our model, we conceptualize a corresponding visual analytics~(VA) design for analytical reasoning in legal case analysis.
Our concept accounts for the inherent complexity of graph-based tours by utilizing aggregated graph nodes and supporting navigation with a semantic lens on the tour's subgraph.
Connecting legal documents through semantic relationships, the concept supports essential tasks of legal case analysis.
% This design acts as a lens on the tour's sub-graph, similar to a fisheye view.
% It enables the task subject to analytical reasoning, including searching and schematizing the data, hypothesizing about its interpretation, and presenting the findings from the analysis.
% For this, the design enables direct search and navigation via semantic relationships, provides semantic context, and offers visual analytical provenance.
We evaluated \framework{} and our conceptualized VA design with six domain experts from law.
They suggest that graph-based tours better support their analytical reasoning than sequences.
Our work opens research opportunities for such tours to support analytical reasoning in law and other knowledge-centric domains.

% We implemented the framework in a Visual Analytics system for exploring legal document collections and evaluated it in a quantitative user study with sixteen legal scholars.
% The results indicate significant reductions in mental workload and measurable improvements in task performance for interpretive work.
% Together, these findings highlight that our framework supports adaptive, semantically informed, and user-driven analytical reasoning.
% Our work lays the foundation for cognitively aligned visual exploration across knowledge-centric domains and has a practical impact on the legal domain.
% TODO: we should point to our future papers here!
% Future work should investigate mixed-initiative guidance that combines algorithmic suggestions with expert-driven exploration while preserving human agency.

%%%%%%%%%%%%%%%%%%%%%%%%%%%%%%%%%%%%%%%%%%%%%%%%%%%%%%%%%%%%%%%%%%%%%%%%%%%%%%%%%%%%%%%%%%%%%%%%

\section*{Acknowledgments}
This work has been partially funded by the Federal Ministry of Research, Technology and Space (BMFTR) in REGAiT (13N17427) and under Germany’s Excellence Strategy - EXC 2117 - 422037984.
We acknowledge the use of Grammarly and GPT-5 solely for grammar correction and language polishing.

%%%%%%%%%%%%%%%%%%%%%%%%%%%%%%%%%%%%%%%%%%%%%%%%%%%%%%%%%%%%%%%%%%%%%%%%%%%%%%%%%%%%%%%%%%%%%%%%

\printbibliography
% \bibliographystyle{eg-alpha-doi} 
% \bibliography{Literature}

%%%%%%%%%%%%%%%%%%%%%%%%%%%%%%%%%%%%%%%%%%%%%%%%%%%%%%%%%%%%%%%%%%%%%%%%%%%%%%%%%%%%%%%%%%%%%%%%

\clearpage
\appendix

\section{Preliminary Evaluation}
\label{appendix:preliminary-evaluation}
As part of our preliminary evaluation, we introduce the participants to our Visual Analytics~(VA) design concept based on \framework{}.
We presented different visualizations for the data view to accompany the semantic view and gauge preferences among the participants.

\subsection{Data View}
We presented the following four visualizations for the data view:

\begin{figure}[!h]
    \centering

    \includegraphics[width=\linewidth]{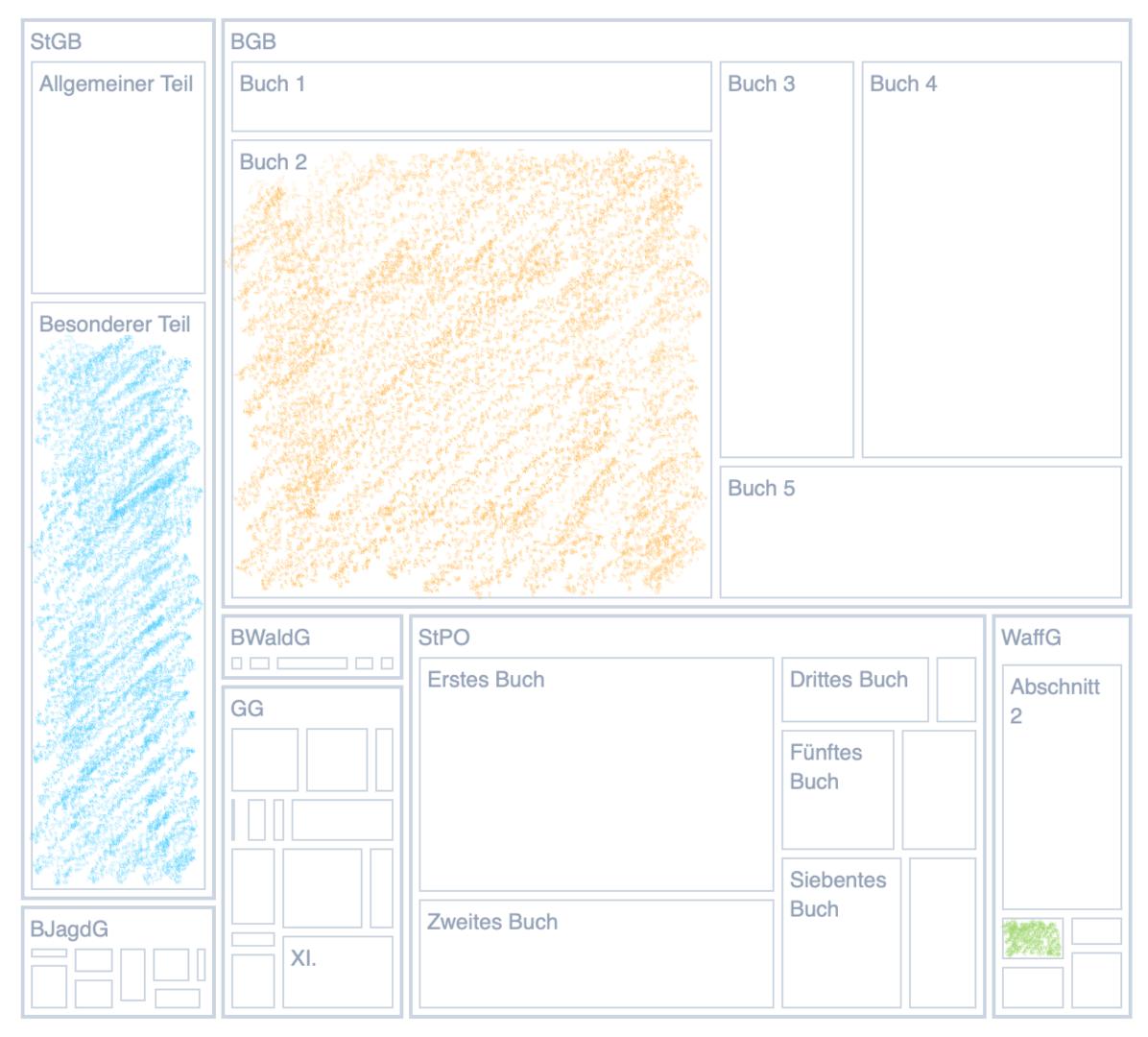}
    
    \caption{Variant 1 of the data view visualizations, resembling a tree map of the structure found in German federal law with colored highlights.}
    \label{appendix:preliminary-evaluation:treemap}
\end{figure}

\begin{figure}[!h]
    \centering

    \includegraphics[width=\linewidth]{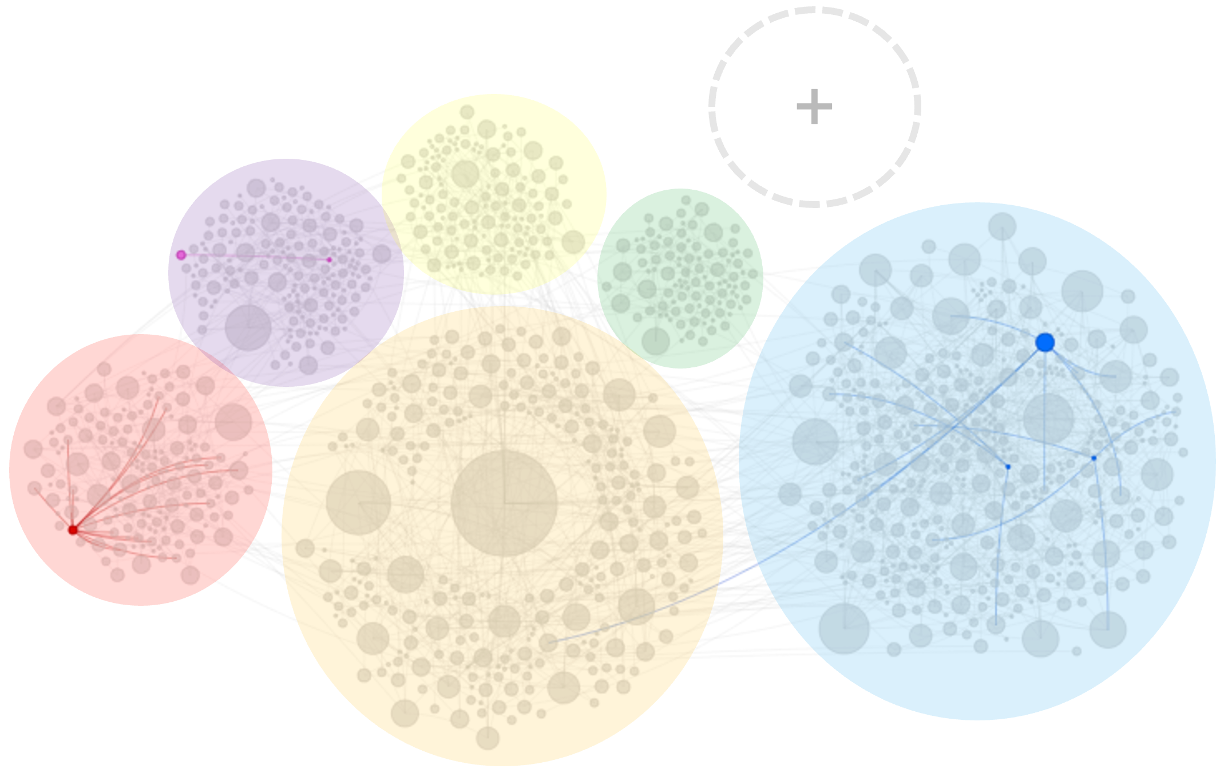}
    
    \caption{Variant 2 of the data view visualizations, resembling a semantic map of legal documents with clusters and colored highlights. Taken from Fürst et al.~\cite{furstChallengesOpportunitiesVisual2025}.}
    \label{appendix:preliminary-evaluation:semantic-map}
\end{figure}

\begin{figure}[!h]
    \centering

    \includegraphics[width=\linewidth]{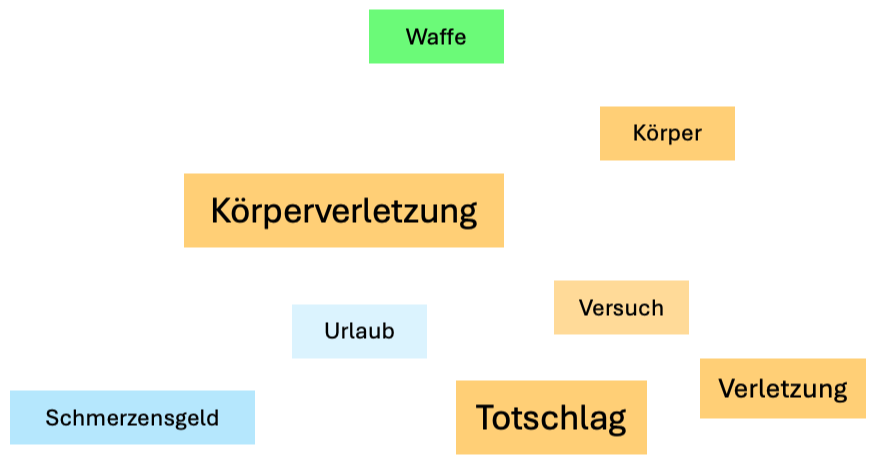}
    
    \caption{Variant 3 of the data view visualizations, resembling a word cloud with color-coded German legal concepts.}
    \label{appendix:preliminary-evaluation:wordcloud-ours}
\end{figure}

\begin{figure}[!h]
    \centering

    \includegraphics[width=0.7\linewidth]{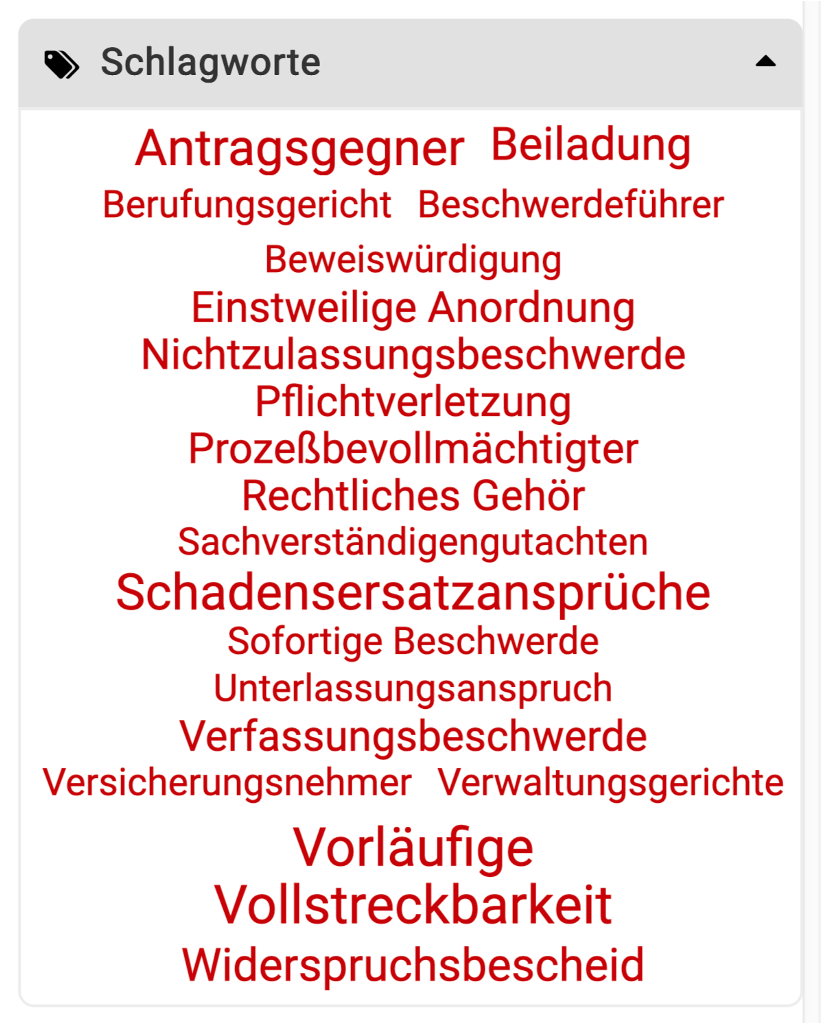}
    
    \caption{Variant 4 of the data view visualizations, resembling a word cloud of German legal concepts taken from beck-online~\cite{verlagc.h.beckohgBeckonline}.}
    \label{appendix:preliminary-evaluation:wordcloud-beck-online}
\end{figure}

\FloatBarrier

\subsection{Semantic View}
We presented the following two visualizations for the semantic view for comparison between the linear and the non-linear tours:

\begin{figure}[!h]
    \centering

    \includegraphics[width=\linewidth]{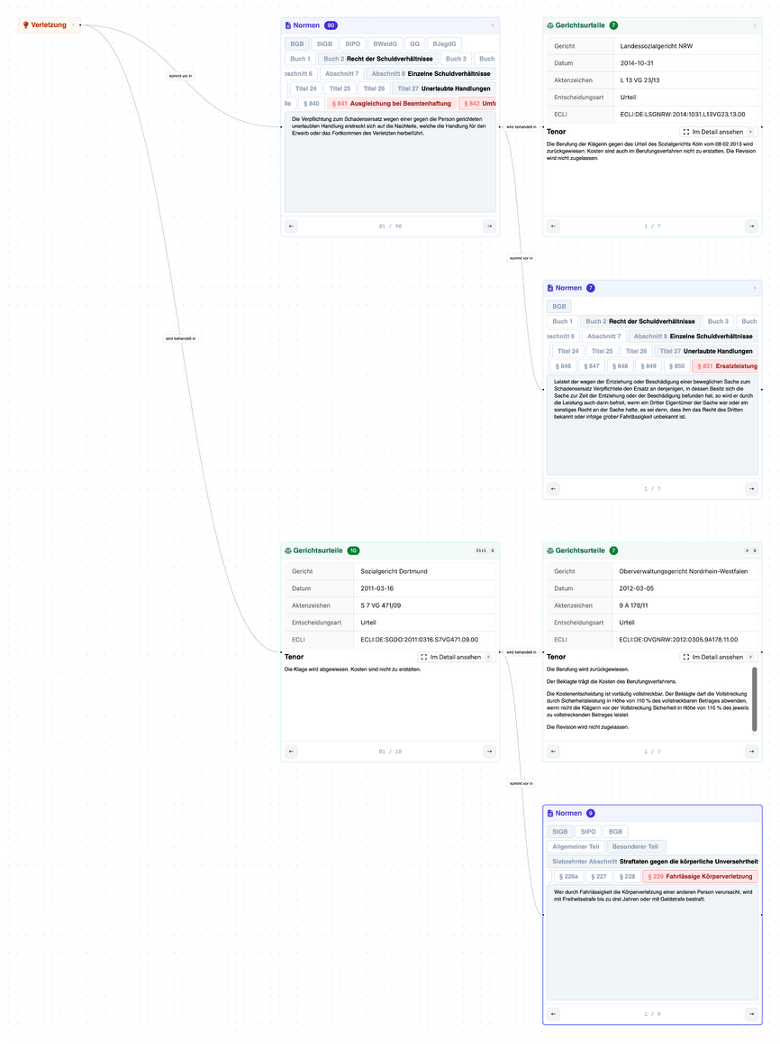}
    
    \caption{Variant 1 of the semantic view visualizations, the non-linear graph of legal documents that is part of our VA concept.}
    \label{appendix:preliminary-evaluation:graph}
\end{figure}

\begin{figure}[!h]
    \centering

    \includegraphics[width=\linewidth]{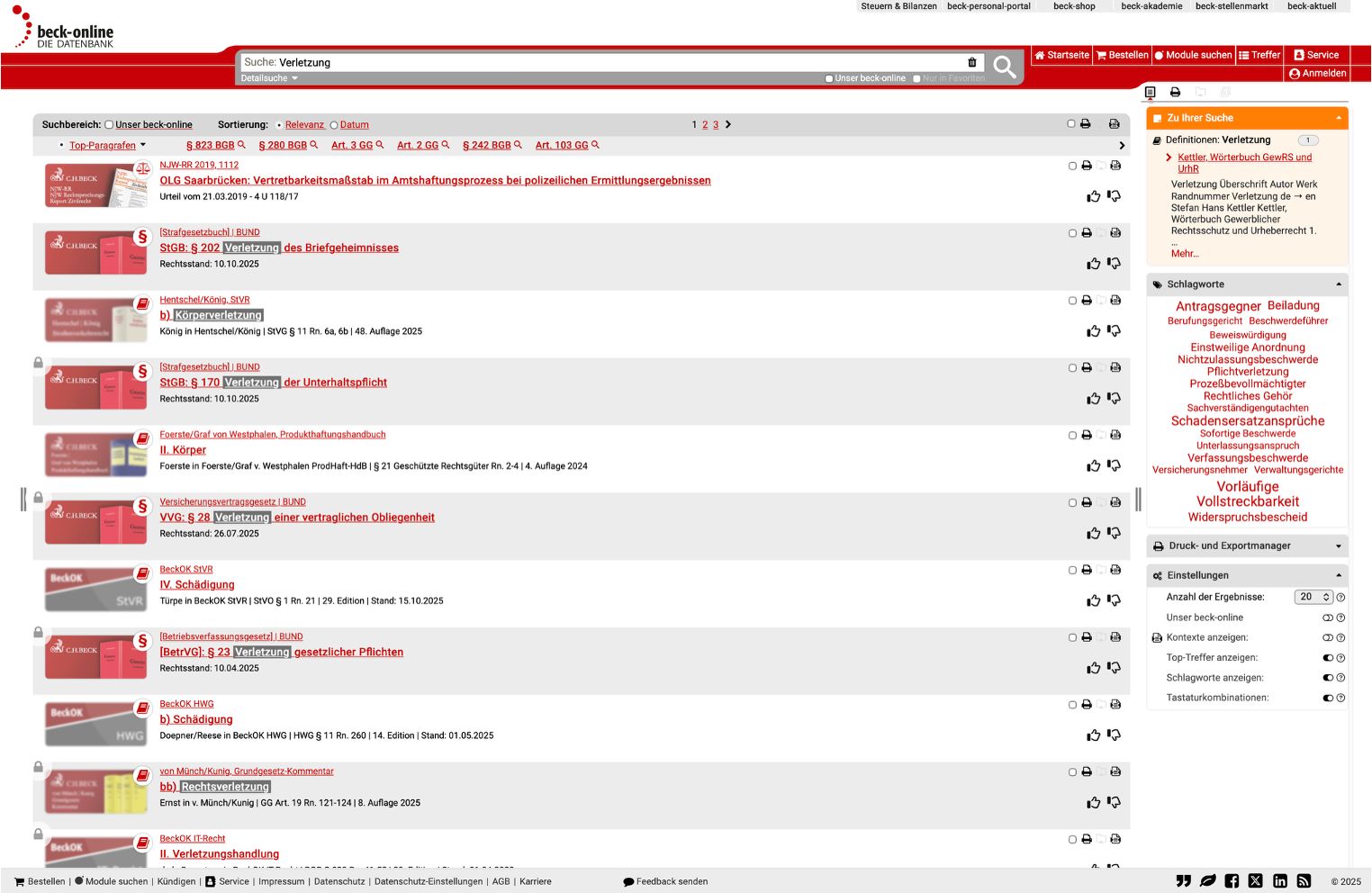}
    
    \caption{Variant 2 of the semantic view visualizations, resembling a linear list of legal documents from beck-online~\cite{verlagc.h.beckohgBeckonline}.}
    \label{appendix:preliminary-evaluation:list}
\end{figure}

\end{document}